\documentclass[aps,
pra,
superscriptaddress,
showpacs,
showkeys,
notitlepage,
twocolumn, 
numbers,
longbibliography,
nofootinbib]{revtex4-2}

\usepackage{lipsum}
\usepackage[utf8]{inputenc}
\usepackage[english]{babel}
\usepackage[T1]{fontenc}
\usepackage{caption}
\usepackage{subcaption}
\usepackage{multirow}
\usepackage{tikz}
\usepackage{amsmath,amssymb,amsthm}
\usepackage{braket}
\usepackage{comment}
\usepackage{hyperref}
\usepackage{multirow}
\usepackage{graphics}
\usepackage{booktabs}
\hypersetup{
     colorlinks=true,
     linkcolor=blue,
     filecolor=blue,
     citecolor=magenta,      
     urlcolor=blue,
}
     
\newcommand{\note}[1]{\textcolor{red}{#1}}

\newcommand{\tr}[1]{\text{Tr}\left({#1}\right)} 

\begin{document}

\title{Non-unitary enhanced transfer efficiency in quantum walk search on complex networks}

\author{Ugo Nzongani}
\email{ugo.nzongani@lis-lab.fr}
\affiliation{Aix-Marseille Université, CNRS, LIS, 13288 Marseille, France} 
\affiliation{Unité de Mathématiques Appliquées, ENSTA, Institut Polytechnique de Paris, 91120 Palaiseau, France}

\author{Andrea Simonetto}
\email{andrea.simonetto@ensta.fr}
\affiliation{Unité de Mathématiques Appliquées, ENSTA, Institut Polytechnique de Paris, 91120 Palaiseau, France}
\affiliation{CEDRIC, Conservatoire National des Arts et Métiers, 75003 Paris, France}

\author{Giuseppe Di Molfetta}
\email{giuseppe.dimolfetta@lis-lab.fr}
\affiliation{Aix-Marseille Université, CNRS, LIS, 13288 Marseille, France}

\begin{abstract}
    The task of finding an element in an unstructured database is known as spatial search and can be expressed as a quantum walk evolution on a graph. In this article, we modify the usual search problem by adding an extra trapping vertex to the graph, which is only connected to the target element. We study the transfer efficiency of the walker to a trapping site, using the search problem as a case study. Thus, our model offers no computational advantage for the search problem, but focuses on information transport in an open environment with a search Hamiltonian. The walker evolution is a mix between classical and quantum walk search dynamics. The balance between unitary and non-unitary dynamics is tuned with a parameter, and we numerically show that depending on the graph topology and the connectivity of the target element, this hybrid approach can outperform a purely classical or quantum evolution for reaching the trapping site. We show that this behavior is only observed in the presence of an extra trapping site, and that depending on the topology and a tunable parameter controlling the strength of the oracle, a hybrid regime composed of 90\% coherent dynamics can lead to either the highest or worst transfer efficiency to the trapping site. We also relate the performance of an hybrid regime to the entropy's decay rate. As the introduction of non-unitary operations may be considered as noise, we interpret this phenomena as a noisy-assisted quantum evolution.
\end{abstract}

\maketitle

\section{Introduction}

Quantum walks are the quantum analog of random walks. They are a coherent model of transportation on graphs and a universal model of quantum computation~\cite{lovett2010universal,Childs_2009}, formulated both in continuous and discrete-time. Continuous-time quantum walks (CTQWs) evolve on the space spanned by the vertices of a graph whose structure is encoded in an Hamiltonian~\cite{PhysRevA.58.915}. As for discrete-time quantum walks (DTQWs), they require the use of a coin to guide the displacements, which enlarges the Hilbert space of the system~\cite{aharonov1993quantum}. Both formalisms are fundamental quantum computing tools as they serve for quantum simulation of fundamental physics~\cite{bepari2022quantum,arnault2016quantum,de2014quantum,di2014quantum,zylberman2022quantum} and quantum field theory~\cite{di2016quantum, eon2023relativistic,sellapillay2022discrete}, quantum information processing and quantum algorithms~\cite{kadian2021quantum,roget2024quantum}. Among their algorithmic applications, few examples are related to optimization problems~\cite{schulz2024guided,PhysRevResearch.2.023302,bennett2021quantum,Slate_2021,campos2021quantum,qu2024experimental,nzongani2025sampled}, quantum state preparation~\cite{Chang_2024,choudhury2024remote,gonzales2025efficient}, machine learning tasks~\cite{shi2024image,dernbach2019quantum,de_Souza_2019, roget2022quantum,sun2025quantumperceptronlearningquantum} or graph related problems~\cite{childs2003exponential,chawla2020discrete,benedetti2023identifying}. Moreover, it has been proved that some DTQWs converge to the Dirac~\cite{Arrighi_2014,di2012discrete,strauch2006relativistic,nzongani2024dirac, di2024quantum} and the Schrödinger~\cite{jolly2023twisted} equations in their continuous limit. 

The task of finding a marked element in an unstructured database is known as spatial search. Naturally, the database is modeled as a graph whose vertices and edges respectively represent its elements and their relationships. The most famous related result is Grover's algorithm~\cite{grover1996fast} which requires $\mathcal{O}(\sqrt{N})$ calls to an oracle to find an element among $N$ in an unstructured database. This algorithm is optimal if the oracle is given as a black box~\cite{zalka1999grover} and was surprisingly shown to be a naturally occurring phenomenon~\cite{roget2020grover}. However, when the inner structure of the oracle is known, a classical quantum-inspired algorithm can potentially solve the search problem exponentially faster by simulating the oracle several times~\cite{PhysRevX.14.041029}. The associated complexity depends on the cost of a single simulation. Moreover, the search problem can be expressed as a quantum walk evolution on graphs, both in discrete~\cite{shenvi2003quantum} and continuous-time~\cite{childs2004spatial}, each resulting in a quadratic speedup on arbitrary graphs~\cite{ambainis2020quadratic,apers2022quadratic}.

Quantum walks in open quantum system can be modeled by Quantum Stochastic Walks (QSWs), which are a generalization of CTRWs and CTQWs~\cite{Whitfield_2010}. They were first introduced as a tool to study the transition between classical and quantum random walks. QSWs have been proposed as an algorithmic tool for several problems including PageRank~\cite{benjamin2024resolving}, decision-making~\cite{martinez2016quantum}, quantum state discrimination~\cite{Dalla_Pozza_2020}, or function approximation and classification~\cite{wang2022implementation}. A discrete-time QSW scheme has also been proposed by Schuhmacher et al.~\cite{schuhmacher2021quantum}. QSWs have been accurately produced experimentally with a three-dimensional photonic quantum chip~\cite{tang2019experimental} and could generally be implemented with the method proposed by Ding et al. for simulation of open quantum systems~\cite{ding2024simulating}. Using the QSWs framework, Caruso has numerically shown that for several graphs, transfer efficiency from an arbitrary vertex to an absorbing vertex, named the sink, is optimal when dynamics is 90\% coherent and 10\% incoherent~\cite{Caruso_2014}. Moreover, Caruso et al. have experimentally implemented a photonic maze from which a single photon must escape, and they recovered the same result: the walker finds his way out faster to the sink when 10\% of the dynamics is non-unitary~\cite{Caruso_2016}. These results suggests that a controlled amount of non-unitary dynamics, which may be interpreted as noise, can improve transfer efficiency from an arbitrary set of vertices to a sink. Lastly, maze solving in open quantum systems has also been studied with QSWs assisted by reinforcement learning~\cite{pozza2021quantum} or with a Grover walk that makes use of sink vertices~\cite{matsuoka2025mathematical}.

In this article, we use continuous-time dynamics to tackle a modified version of the search problem for single marked element. We introduce the Stochastic Quantum Walk Search (SQWS) monitored by a weighted Lindbladian. The unitary evolution is induced by Childs and Goldstone's CTQW search Hamiltonian~\cite{childs2004spatial} where we parametrize the oracle instead of the walk generator, and non-unitary dissipation is designed to implement a CTRW search dynamics. In addition, we use a trapping sink vertex as an extra dissipative tool, which we only connect to the target vertex of the search with an irreversible transition. Although this is a search problem, we do not quantify performance by the time taken to reach the marked element as a function of graph size as often done, but by the transfer efficiency to reach the trapping site. This metric measures both the amount of information present in the sink and the time needed to reach it. It has already been used as a measure of transport quality in previous works \cite{Caruso_2014,pozza2021quantum,Chin_2010,PhysRevA.81.062346}, and verified experimentally \cite{Caruso_2016}. In this work, we are interested in a transport problem, not a computational one, and we use the search problem as a case study. The walker starts from a uniform superposition over the vertices of the graph and has to reach the trapping site by moving through a search-driven dynamics guiding it to the target vertex. We numerically show that a mix of unitary and non-unitary operations can outperform a fully coherent or incoherent dynamics for reaching the sink. The balance between unitary and non-unitary operations is controlled with a tunable mixing parameter. The performance of the hybrid regime depends on the graph topology and the connectivity of the target vertex. We show that a low-noise regime can outperform others regimes only for graphs with low density, high eccentricity and low degree centrality for the target vertex. Moreover, we show that the addition of non-unitary operations leads to an improvement of performance only in the presence of a sink. Lastly, we relate the performance of a mixing regime to the system's entropy decay rate. We interpret the hybrid regime as a noisy-assisted quantum evolution as it contains non-unitary operations, even if does not model realistic hardware noise.

We start by introducing different continuous-time dynamics on graphs and the search problem of a single element in Sec. \ref{sec:prelim}. We present our model for the modified version of the search problem and our results in Sec. \ref{sec:results}.

\section{Preliminaries}\label{sec:prelim}

\subsection{Continuous-time dynamics}

\subsubsection{Random Walks}
Continuous-time Random Walks (CTRWs) describe the evolution of a single walker on a network modeled by a graph $G=(V,E)$ where $V$ and $E$ are respectively a set of vertices and edges. An unweighted graph is fully described by its adjacency matrix:
\begin{equation}
    A_{ij} = 
    \begin{cases}
        1 \text{ if } (i,j) \in E, \\
        0 \text{ otherwise}.
    \end{cases}
\end{equation}
The CTRW is a Markov process whose rate matrix is the Laplacian $L=D-A$ of the graph, with $D$ a matrix whose diagonal entries are the degree of each vertex. The state of the walker is described by a probability distribution $\Vec{p}$ which is a map from $V$ to probabilities. The time evolution of the walker is:
\begin{equation}\label{eq:random_walk}
    \frac{d}{dt}\Vec{p}(t)=-L\Vec{p}(t).
\end{equation}
As the column of $L$ sum to zero, an initially normalized probability distribution remains valid under the evolution induced by Eq. \eqref{eq:random_walk}.

\subsubsection{Quantum Walks}

Continuous-time quantum walks are a coherent model of transportation over complex networks~\cite{mulken2011continuous}. The walker is represented with a quantum state $\ket{\psi}$ that evolves in an Hilbert space $\mathcal{H}$ spanned by the vertices of the graph. Therefore, the set of vertices $V$ form an orthonormal basis of $\mathcal{H}$. The time evolution of the walker is given by the Schrödinger equation and the Hamiltonian encodes the structure of the graph:
\begin{equation}
    i\hbar\frac{d}{dt}\ket{\psi(t)} = H\ket{\psi(t)}.
\end{equation}
Throughout, we work in units in which $\hbar=1$. As the evolution is unitary, the Hamiltonian $H$ has to be Hermitian, making the underlying graph undirected, which is not the case for CTRWs as their underlying graph can have both directed and undirected edges.

\subsubsection{Quantum Stochastic Walks}

The transition from the classical to the quantum regime can be studied with the QSW framework that enables to interpolate between coherent and incoherent dynamics~\cite{Whitfield_2010}. The state of the walker is described by a density matrix $\rho=\ket{\psi}\bra{\psi}$ as the evolution is composed of both unitary and non-unitary operations. The QSWs evolution is driven by a weighted Lindblad master equation with $\omega\in [0,1]$:
\begin{equation}\label{eq:lindblad}
    \small{
    \begin{split}
    \frac{d}{dt}\rho(t) &= -i(1\!-\!\omega)[H,\rho(t)] \!+\! \omega\!\sum_{ij}\!\Big(\!\mathcal{L}_{ij}\rho(t)\mathcal{L}_{ij}^\dagger\!-\!\frac{1}{2}\!\{\!\mathcal{L}_{ij}^\dagger \mathcal{L}_{ij},\rho(t)\!\}\!\Big) \\
    &= (1-\omega)\mathcal{U}_H[\rho(t)] + \omega \mathcal{D_L}[\rho(t)],
    \end{split}
    }
\end{equation}
where $[\alpha,\beta]=\alpha\beta-\beta\alpha$ and $\{\alpha,\beta\}=\alpha\beta+\beta\alpha$ are the commutator and anti-commutator of operators $\alpha$ and $\beta$. The unitary and non-unitary dynamics are respectively encapsulated by $\mathcal{U}_H$ and $\mathcal{D_L}$, and the parameter $\omega$ enables to interpolate between them. A wise choice of Lindblad jump operators $\mathcal{L}_{ij}$ can describe a CTRW evolution~\cite{Whitfield_2010}. Therefore, for well-defined Lindblad jump operators, the CTQW is recovered for $\omega=0$ and the CTRW for $\omega=1$. A linear combination of the two is obtained
for other values of $\omega$, leading to a mix between coherent and incoherent dynamics for the walker. It was shown that the introduction of an extra sink vertex in Eq. \eqref{eq:lindblad}, with $H=A$ and $\mathcal{L}_{ij}=A_{ij}/D_{ij}\ket{i}\bra{j}$, leads to an optimal transfer from a set of vertices to this sink vertex when $\omega=0.1$ for several graphs~\cite{Caruso_2014}.

\subsection{Spatial Search}

\subsubsection{Random Walk Search}

Spatial search expressed as a CTRW evolution consists of a walker spreading over the vertices of a graph until it reaches the marked element $m$. The marked vertex is an absorbing vertex, meaning that the probability of leaving it is zero. Therefore, the Laplacian has to be modified into an absorbing Laplacian $\hat{L}=\frac{D-\hat{A}}{||D-\hat{A}||_2}$ where $\hat{A}$ is the adjacency matrix whose $m$-th column is the $m$-th canonical basis vector of $\mathbb{R}^{|V|}$~\cite{Wong_2022}. The initial state is the uniform probability distribution over the vertices of the graph:
\begin{equation}
    \Vec{p}(0)=\frac{1}{|V|}\sum_{v\in V}\Vec{e}_v,
\end{equation}
where $\{\Vec{e}_v|v\in V\}$ is the set of canonical basis vectors of $\mathbb{R}^{|V|}$.

\subsubsection{Quantum Walk Search}

An optimal quantum walk search algorithm on arbitrary graphs has been introduced by Apers et al.~\cite{apers2022quadratic}. However, their algorithm implies an expansion of system's Hilbert space as their evolution requires the preparation of an auxiliary Gaussian state. As we do not want to increase the size of the Hilbert space, we use the Hamiltonian introduced by Childs and Goldstone that allows an evolution in the space spanned by the vertices of the graph, even if this framework is not optimal for all graphs~\cite{childs2004spatial}. Thus, the search Hamiltonian is defined as:
\begin{equation}\label{eq:search_hamiltonian}
    H_{m,\gamma} =  L - \gamma\ket{m}\bra{m},
\end{equation}
where $\gamma\in \mathbb{R}^+$ determines the strength of the oracle and $L$ is the graph Laplacian\footnote{Note that in the numerical simulations we have normalized the Laplacian so that its eigenvalues are between 0 and 1, thus $L=I-L/\lambda_{\max}$ where $\lambda_{\max}$ is the highest eigenvalue of $L$.}. Note that in the literature the search Hamiltonian is usually defined as $H_{\text{search}}=\gamma^{-1}H_{m,\gamma}$ and can be recovered from $H_{m,\gamma}$ with proper time rescaling $t=t_{\text{search}}\gamma^{-1}$. However, we parametrize the oracle instead of the Laplacian to control the strength of the oracle instead of the hopping rate of the walk. The initial state is the uniform superposition over the vertices of the graph:
\begin{equation}
    \ket{\psi(0)} = \frac{1}{\sqrt{|V|}}\sum_{v\in V}\ket{v}.
\end{equation}
The heart of the CTQWs spatial search algorithm is to find the minimal value of $t$ and the optimal value of $\gamma$ to maximize the success probability $|\bra{m}e^{-itH_{m,\gamma}}\ket{\psi(0)}|^2$ of finding the marked element. This algorithm was first shown to offer a quadratic speedup over its classical counterparts for the complete graph, the hypercube and the $d$-dimensional periodic lattice for $d>4$~\cite{childs2004spatial}, and to be optimal for a wide family of graphs by Chakraborty et al.~\cite{chakraborty2016spatial}. Lastly, it was shown that quantum walk search performances in this framework can be predicted if certain conditions on the spectral properties of the Hamiltonian driving the walk are met~\cite{chakraborty2020optimality}.

\section{Results}\label{sec:results}

\subsection{Stochastic Quantum Walk Search}

\subsubsection{Model}

\begin{figure}
\begin{center}
\begin{tikzpicture}[scale=0.1]
\tikzstyle{every node}+=[inner sep=0pt]
\draw [black] (37,-20.3) circle (3);
\draw (37,-20.3) node {$m$};
\draw [black] (26.5,-26.6) circle (3);
\draw [black] (46.7,-36.7) circle (3);
\draw [black] (26.5,-36.7) circle (3);
\draw [red] (51.2,-13.6) circle (3);
\draw (51.2,-13.6) node {\textcolor{red}{$\phi$}};
\draw [red] (51.2,-13.6) circle (2.4);
\draw [black] (37,-42.9) circle (3);
\draw [black] (46.7,-26.6) circle (3);
\draw [black] (29.07,-25.06) -- (34.43,-21.84);
\fill [black] (34.43,-21.84) -- (33.48,-21.83) -- (34,-22.68);
\draw [black] (34.43,-21.84) -- (29.07,-25.06);
\fill [black] (29.07,-25.06) -- (30.02,-25.07) -- (29.5,-24.22);
\draw [red] (39.71,-19.02) -- (48.49,-14.88);
\fill [red] (48.49,-14.88) -- (47.55,-14.77) -- (47.98,-15.67);
\draw (45.28,-17.46) node [below] {$\Gamma$};
\draw [black] (26.5,-29.6) -- (26.5,-33.7);
\fill [black] (26.5,-33.7) -- (27,-32.9) -- (26,-32.9);
\draw [black] (26.5,-33.7) -- (26.5,-29.6);
\fill [black] (26.5,-29.6) -- (26,-30.4) -- (27,-30.4);
\draw [black] (34.42,-41.37) -- (29.08,-38.23);
\fill [black] (29.08,-38.23) -- (29.52,-39.06) -- (30.03,-38.2);
\draw [black] (29.08,-38.23) -- (34.42,-41.37);
\fill [black] (34.42,-41.37) -- (33.98,-40.54) -- (33.47,-41.4);
\draw [black] (39.53,-41.28) -- (44.17,-38.32);
\fill [black] (44.17,-38.32) -- (43.23,-38.33) -- (43.77,-39.17);
\draw [black] (44.17,-38.32) -- (39.53,-41.28);
\fill [black] (39.53,-41.28) -- (40.47,-41.27) -- (39.93,-40.43);
\draw [black] (46.7,-29.6) -- (46.7,-33.7);
\fill [black] (46.7,-33.7) -- (47.2,-32.9) -- (46.2,-32.9);
\draw [black] (46.7,-33.7) -- (46.7,-29.6);
\fill [black] (46.7,-29.6) -- (46.2,-30.4) -- (47.2,-30.4);
\draw [black] (44.18,-24.97) -- (39.52,-21.93);
\fill [black] (39.52,-21.93) -- (39.91,-22.79) -- (40.46,-21.95);
\draw [black] (39.52,-21.93) -- (44.18,-24.97);
\fill [black] (44.18,-24.97) -- (43.79,-24.11) -- (43.24,-24.95);
\end{tikzpicture}
\end{center}
\caption{Modified cycle graph $C_6$ on which we add the extra sink vertex $\phi$ connected to the target vertex $m$. The irreversible transition from $m$ to $\phi$ is weighted with the sink rate $\Gamma$. The classical and quantum walk search dynamics only act on the vertices of the initial graph, not on the sink vertex. The search dynamics guides the walker to the target vertex $m$ for it to reach the sink $\phi$, and “escape” the graph.}
\label{fig:graph_example}
\end{figure}
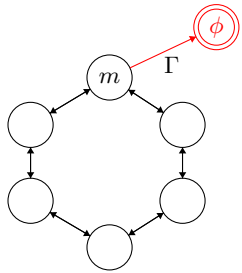

The SQWS is composed of CTQW and CTRW search dynamics with an additional sink. The sink plays a fundamental role in the system, as we later show that the introduction of non-unitary dynamics improves performance only in its presence. We introduce the sink vertex $\phi$ and connect it to the target vertex $m$. The irreversible transition from $m$ to $\phi$ is modeled by the Lindblad jump operator:
\begin{equation}
    \mathcal{L}_{\phi,m}[\rho(t)] = \ket{\phi}\bra{m}\rho(t)\ket{m}\bra{\phi}-\frac{1}{2}\{\ket{m}\bra{m},\rho(t)\}.
\end{equation}
Therefore, the evolution of the SQWS is:
\begin{equation}\label{eq:qsw_search}
    \begin{split}
        \frac{d}{dt}\rho(t) &= (1-\omega)\mathcal{U}_{H_{m,\gamma}}[\rho(t)] + \omega \mathcal{D}_{\mathcal{L}}[\rho(t)] +\Gamma\mathcal{L}_{\phi,m}[\rho(t)] \\
        &= \mathcal{E}[\rho(t)],
    \end{split}
\end{equation}
with $\Gamma\in \mathbb{R}^+$ the sink rate and $\mathcal{L}_{ij}=-\hat{L}_{ij}\ket{i}\bra{j}$. Thus, the coherent and incoherent dynamics respectively produce a quantum and classical random walks search. The initial state is the uniform superposition over the vertices of the graph (excluding the sink vertex) $\rho(0)=\ket{\psi(0)}\bra{\psi(0)}$\footnote{According to our definitions, $\rho(0)$ evolves in a $|V|$-dimensional Liouville space $\mathcal{B}[\mathcal{H}]$. However, as we add the sink vertex $\phi$ to the graph, its dimension should be $|V|+1$. By abuse of notation, we assume that all the operators defined in Eq. \eqref{eq:qsw_search} and $\rho(0)$ act on the ($|V|+1$)-dimensional Liouville space spanned by the vertices of the graph and the additional sink vertex $\phi$.}. As an illustration, we show in Fig. \ref{fig:graph_example} a graph with the extra sink vertex $\phi$ connected to the target vertex $m$.

In this article, we study the transfer efficiency from an uniform superposition over the vertices to the sink. The walker is guided to the target vertex with a search Hamiltonian, and non-unitary operations designed to implement a CTRW-search, which we interpret as noise. Once on the target vertex, the walker may jump to the sink and remains trapped inside. Starting from a uniform superposition over the vertices, we measure performance with transfer efficiency to the sink (connected only to the marked vertex). The transfer efficiency is defined as \cite{Caruso_2014,pozza2021quantum,Chin_2010,PhysRevA.81.062346,Caruso_2016}:
\begin{equation}\label{eq:efficiency}
    E(\omega,\gamma,t) = \frac{1}{t}\int_{0}^{t} \tr{\rho(\tau)\ket{\phi}\bra{\phi}} \,\text{d}\tau.
\end{equation}

We normalize Eq. \eqref{eq:efficiency} so that $E(\omega,\gamma,t)=1$ corresponds to a total instantaneous transfer of the walker to the sink vertex\footnote{Note that instantaneous transfer of the walker to the sink vertex is not realistic. The value $E(\omega,\gamma,t)=1$ is just used as an ideal upper bound to evaluate transfer quality.}, and $E(\omega,\gamma,t)=0$ that its probability of presence on the sink is zero. As, Eq. \eqref{eq:efficiency} considers both success probability and its associated evolution time, it is our unique performance metric. Moreover, we have also executed the SQWS with no sink and show that its presence is mandatory for the hybrid dynamics to beat a fully classical or quantum dynamics. In the absence of sink, the introduction of non-unitary dynamics always reduces performance as we show in Appendix~\ref{app:no_sink}.

\subsubsection{Numerical results}

\begin{figure*}
    \centering
    \includegraphics[width=0.3\textwidth]{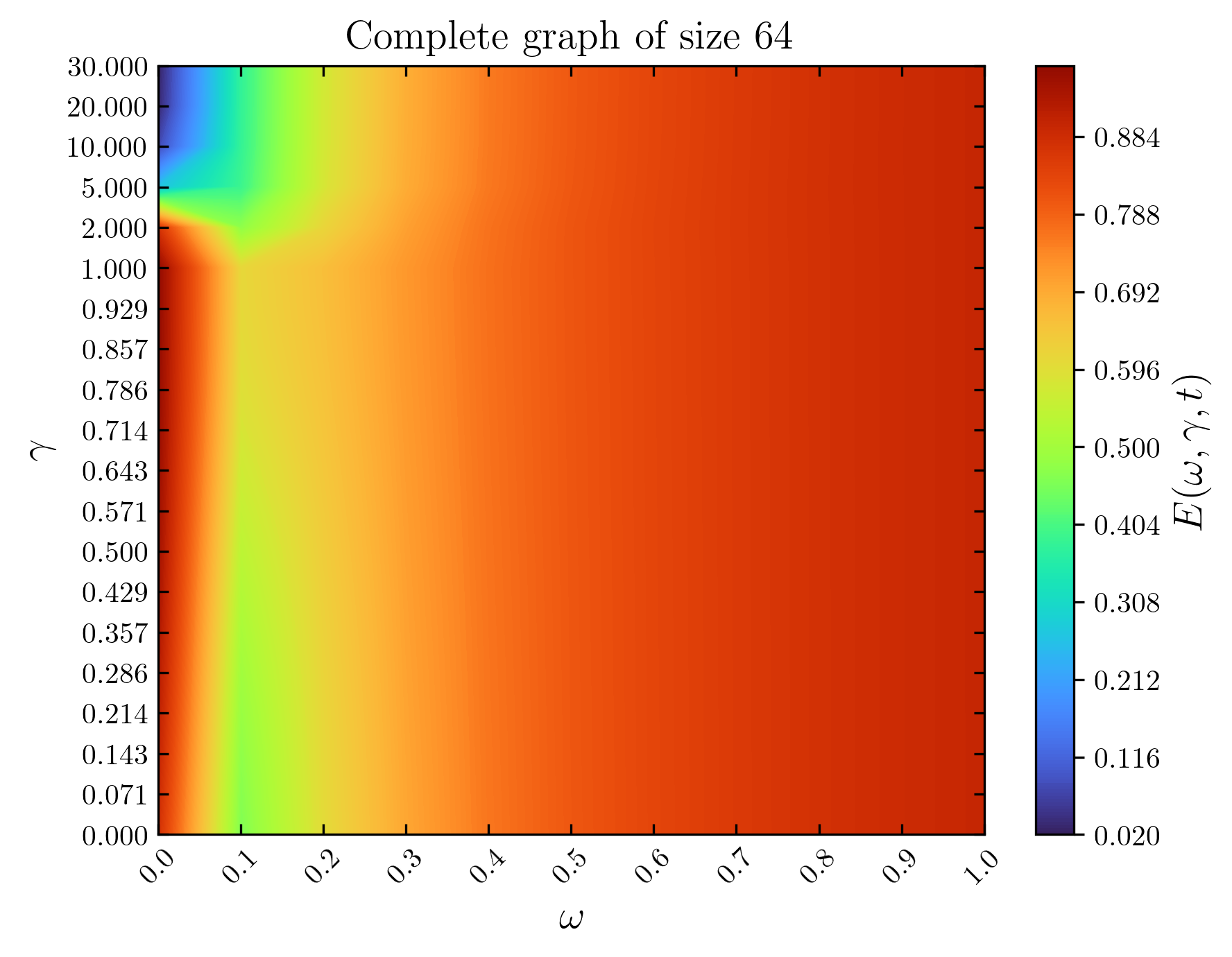}
    \includegraphics[width=0.3\textwidth]{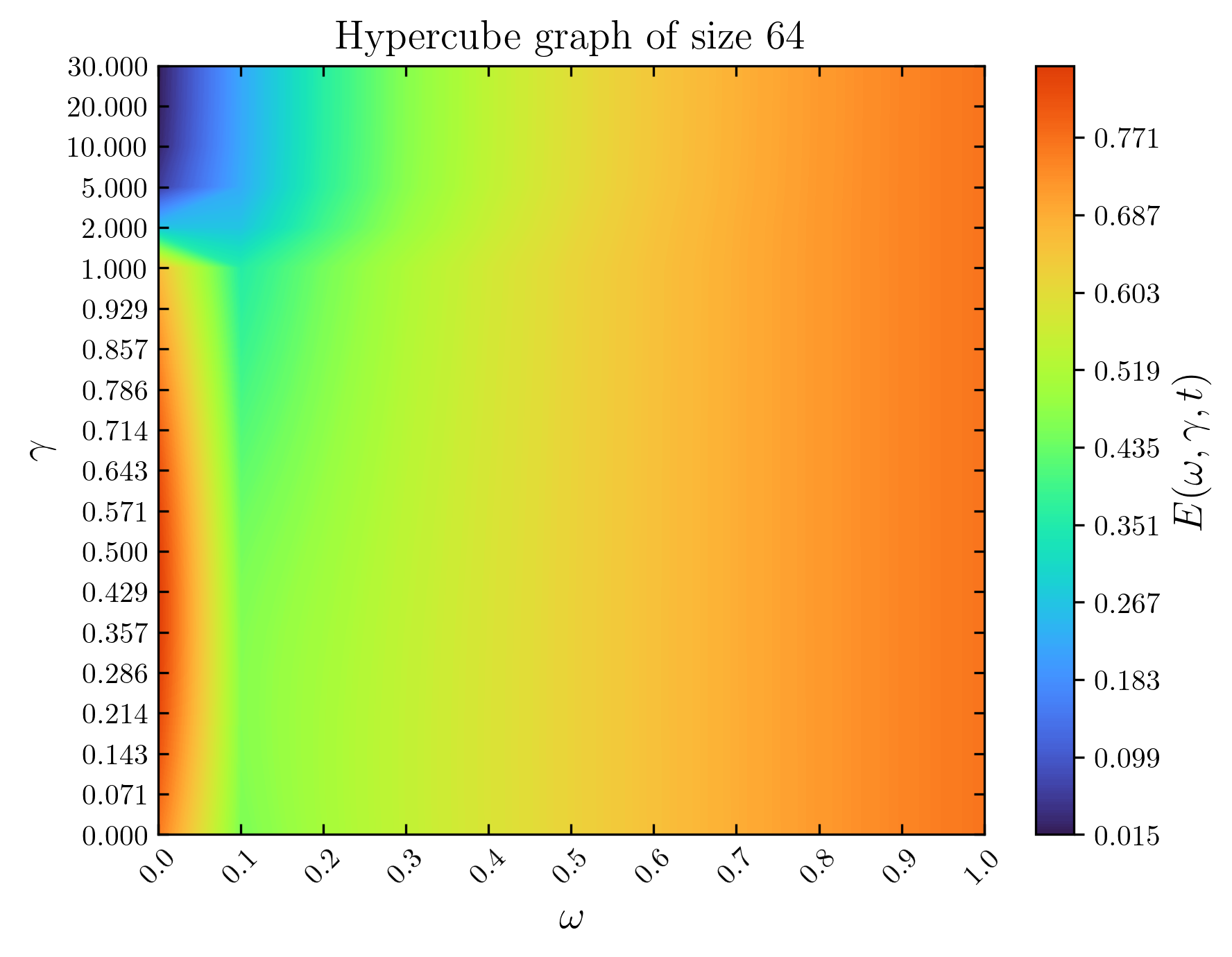}
    \includegraphics[width=0.3\textwidth]{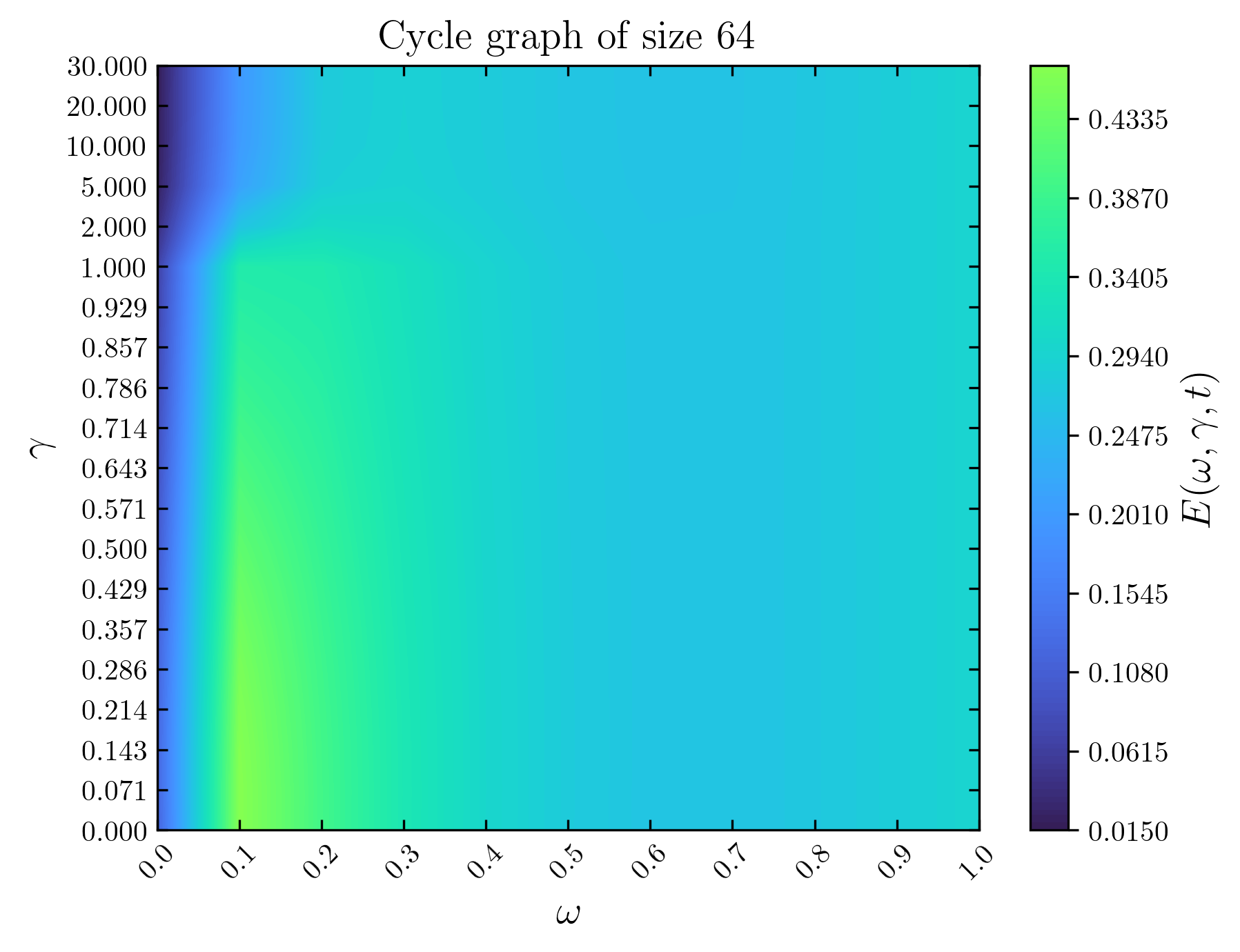}
    \includegraphics[width=0.3\textwidth]{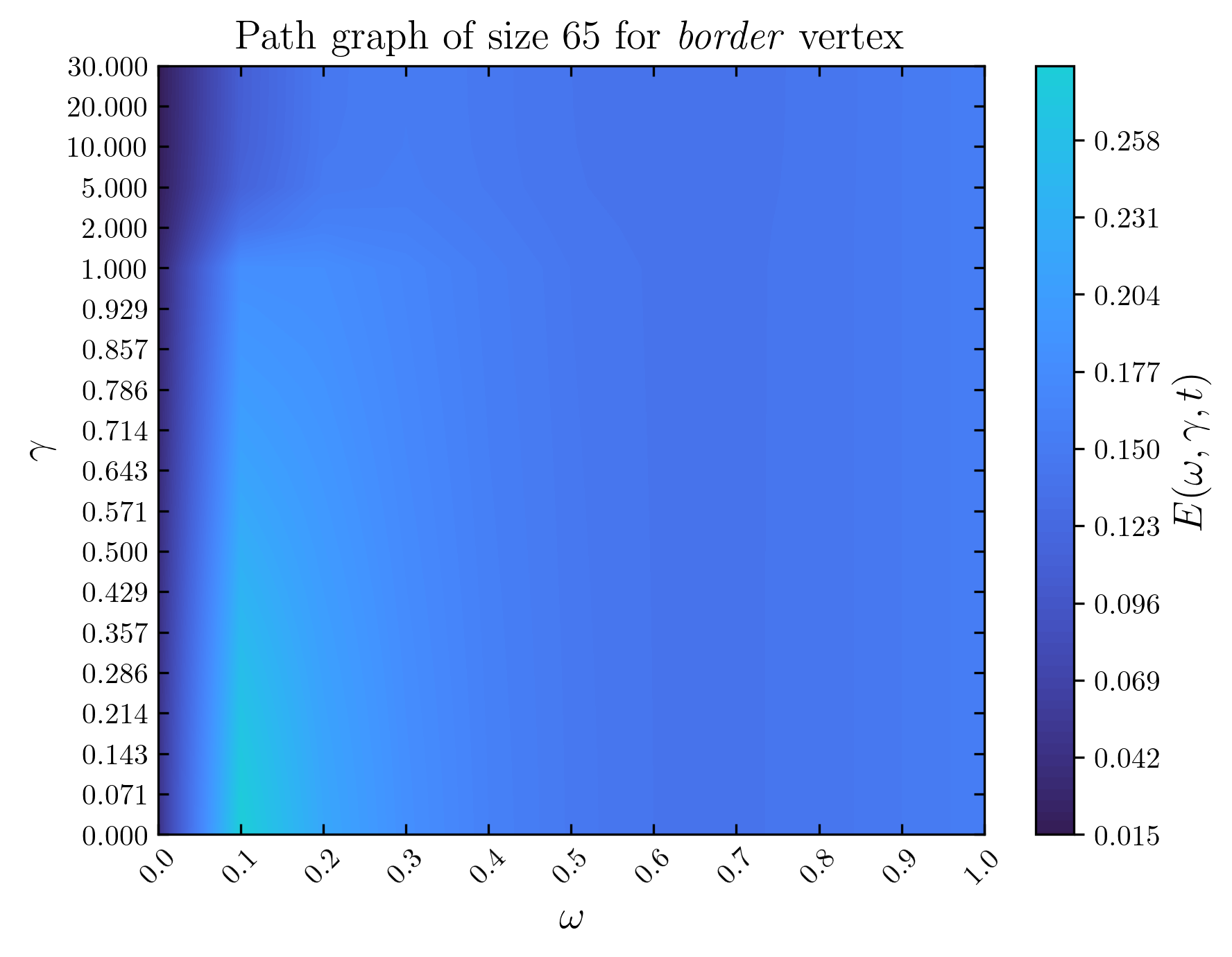}
    \includegraphics[width=0.3\textwidth]{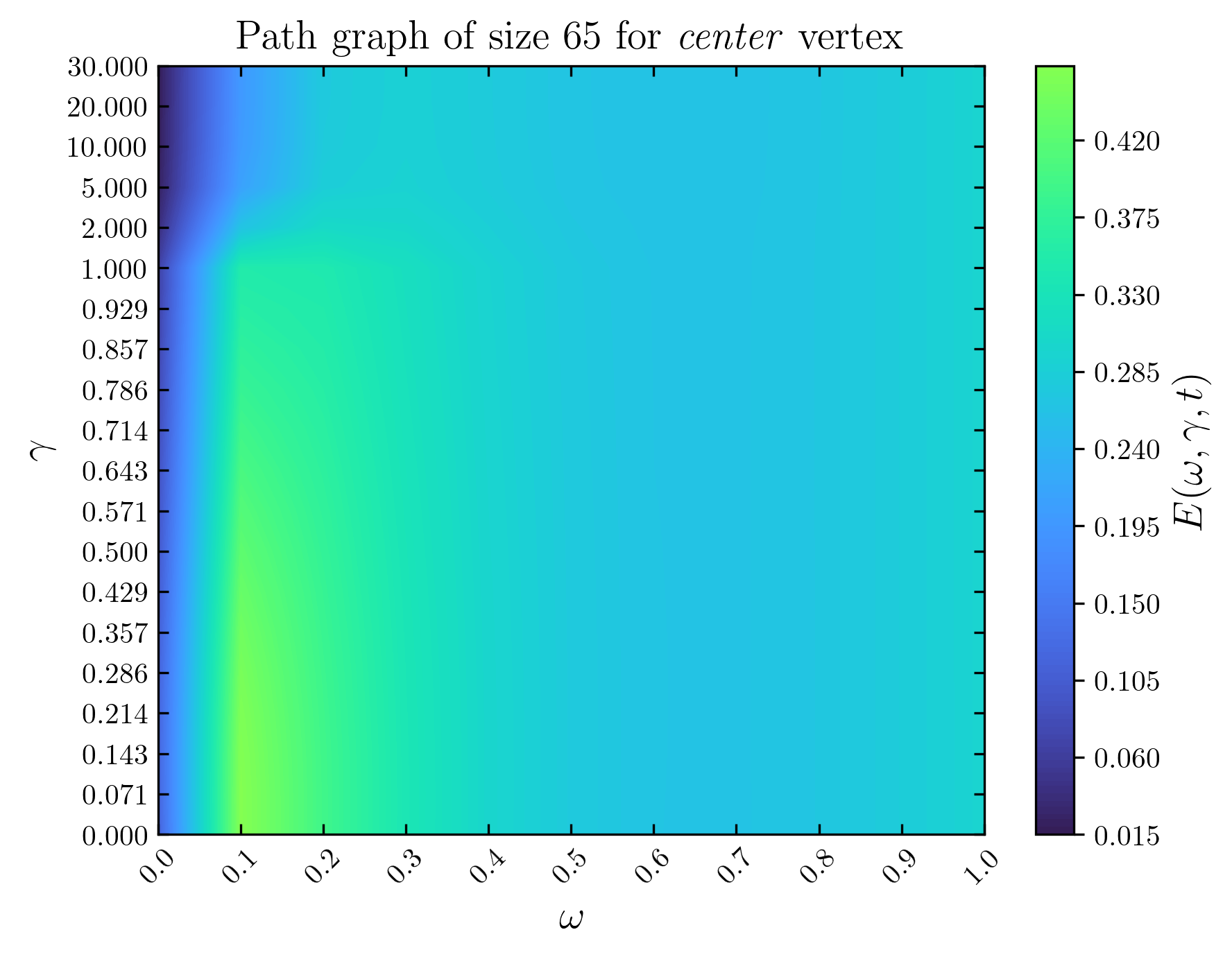}
    \includegraphics[width=0.3\textwidth]{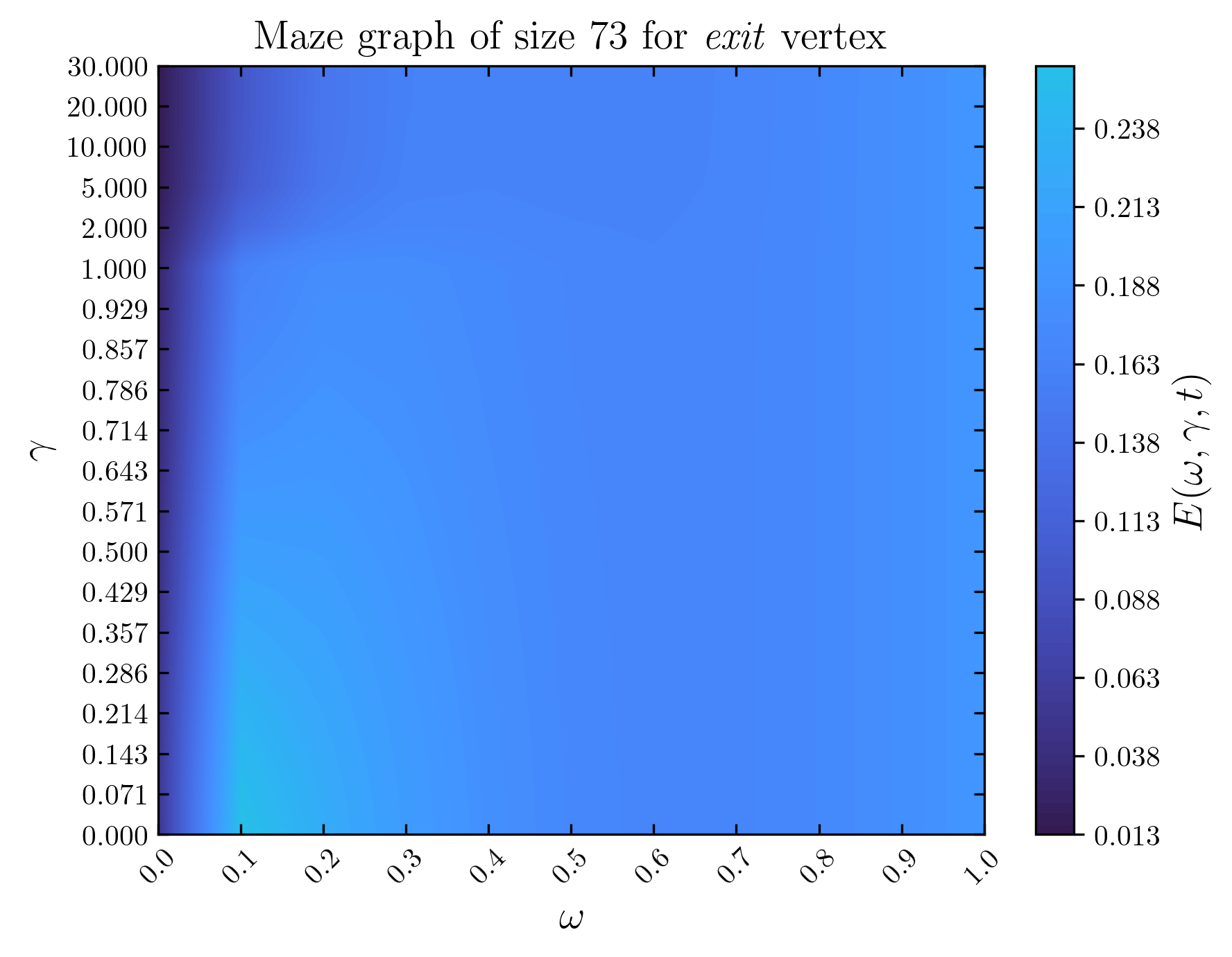}
    \includegraphics[width=0.3\textwidth]{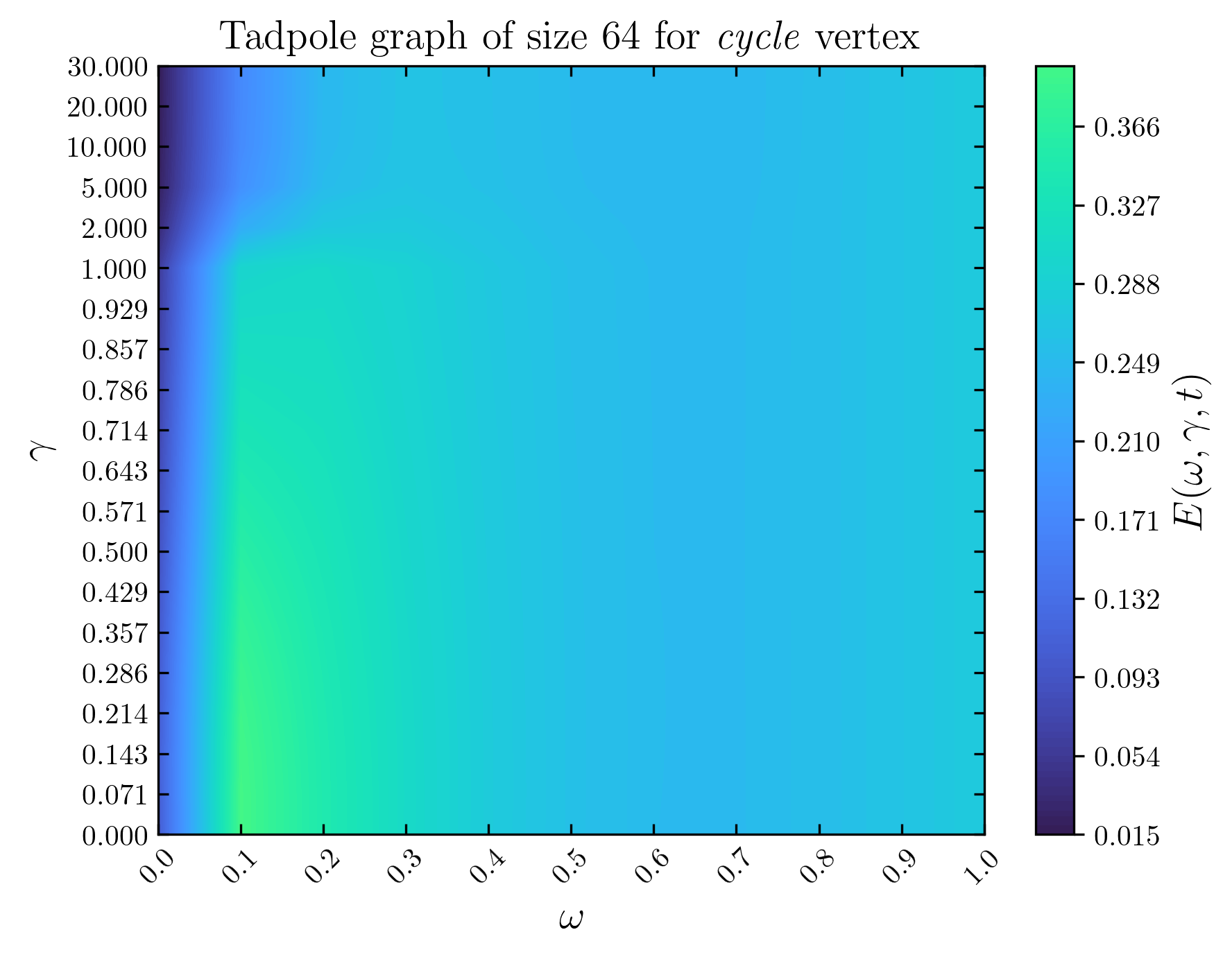}
    \includegraphics[width=0.3\textwidth]{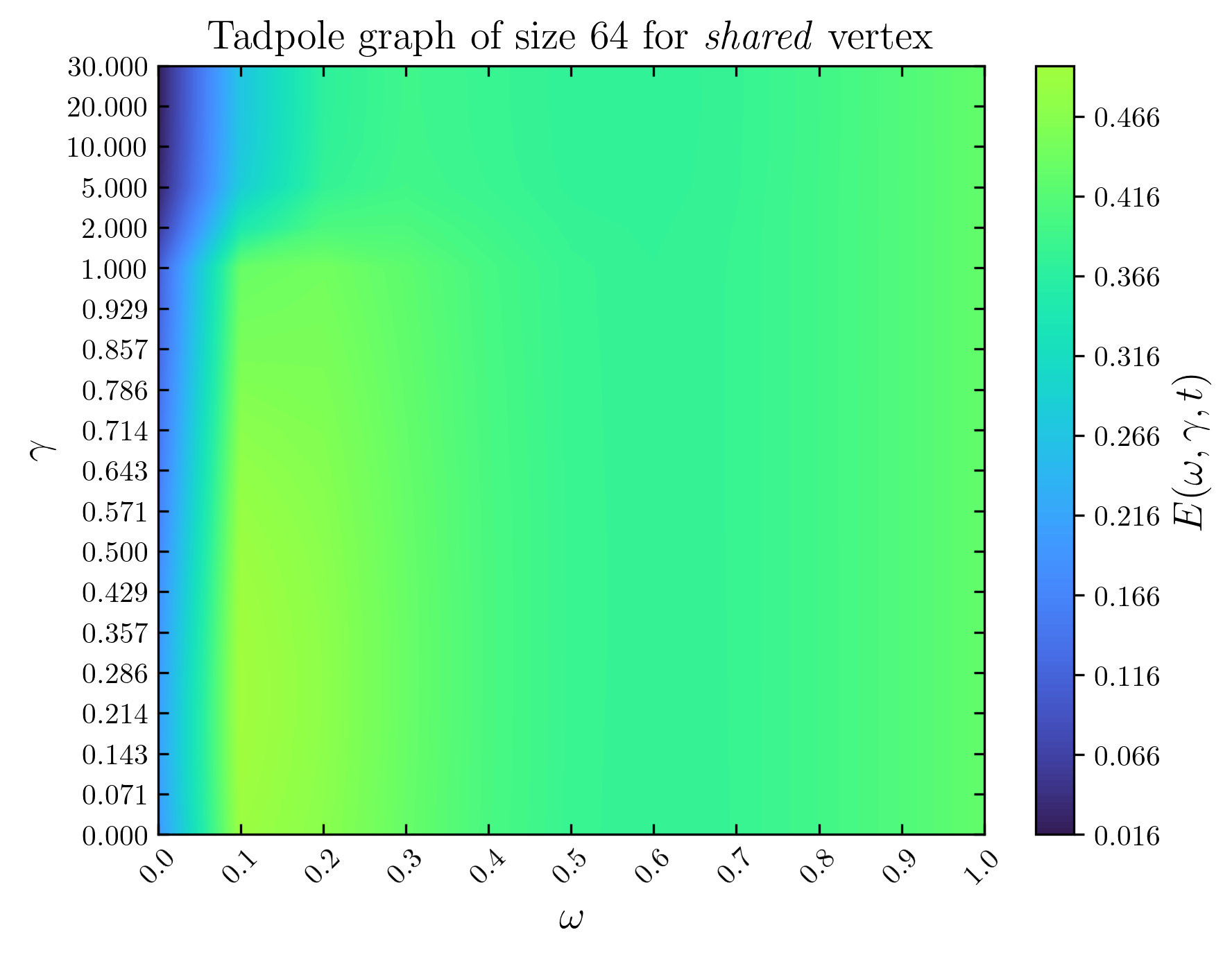}
    \includegraphics[width=0.3\textwidth]{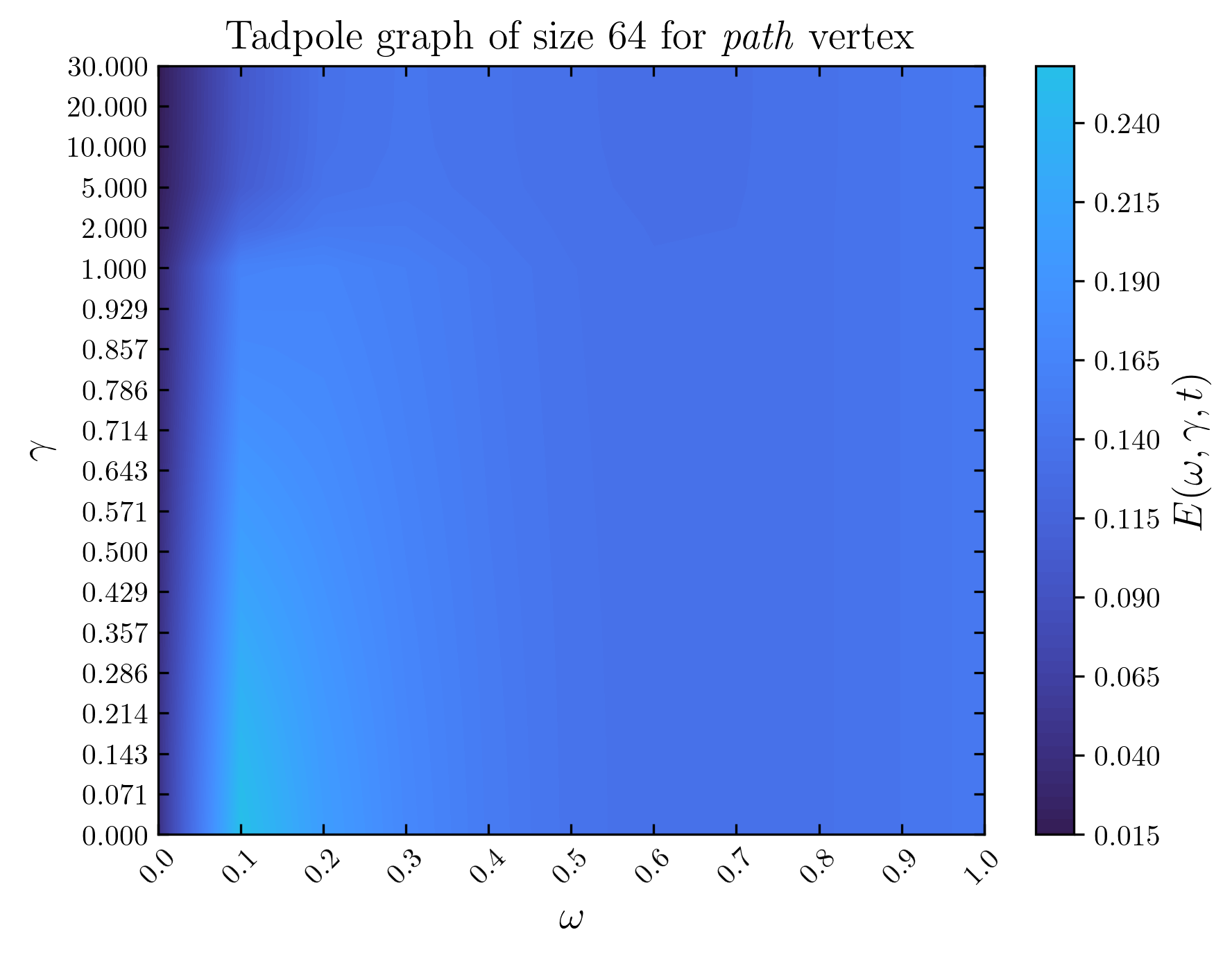}
    \caption{Stochastic Quantum Walk Search (SQWS) performances with different values of $\gamma\in [0,30]$ on graphs of 64 vertices, namely, the complete graph ($K_{64}$), the 6-dimensional hypercube graph ($Q_6$), the cycle graph ($C_{64}$), the path graph ($P_{65}$), a maze graph ($M_{73}$) and the tadpole graph ($T_{32,32}$). The time evolution is set to $t=640$ units of time. The quantum walk dynamics is recovered for $\omega=0$, the classical random walk for $\omega=1$, and a linear combination of the two when $\omega\in]0,1[$.}
    \label{fig:qsws}
\end{figure*}

We run the SQWS on instances of different graph families and show the results in Fig. \ref{fig:qsws}. We set the time to be linear with the size of the graph as we do not measure performance with computational complexity of reaching the marked element, but with transfer efficiency to the sink. Thus, the value of the maximum evolution time is not as important as for the study of the usual quantum search problem where the optimal performance leads to a quadratic speedup over classical methods. For this reason, we select a large time evolution $t=10|V|$. Moreover, we set the sink rate to $\Gamma=1$. For each graph, we run the SQWS for different fixed values of $\gamma$, which controls the strength of the oracle in Eq. \eqref{eq:search_hamiltonian}. For $\gamma=0$, the coherent evolution only induces a free quantum walk exploration, and setting $\gamma>0$ impacts the strength of the oracle marking the target vertex. In the following, we use graphs of size $N\in [32,81]$.

We choose fixed values of $\gamma\in[0,30]$, taking low and high values to see the impact of the value of this parameter on the transfer efficiency to the sink vertex. We present the main results in Fig. \ref{fig:qsws} which illustrates the execution of the SQWS on different graphs, namely, the complete graph $K_{64}$, the 6-dimensional hypercube $Q_6$, the cycle graph $C_{64}$, the path graph $P_{65}$, a maze graph\footnote{Maze generation can be easily done using Depth-First Search (DFS) on a grid. Once the maze is created, each cell is considered as a vertex, and two vertices are adjacent if there is no wall between their respective cells.} ($M_{73}$) and the tadpole graph $T_{32,32}$. The tadpole $T_{M,N}$ graph is the fusion between a cycle graph of size $M$ with a path of size $N$, we display an instance of this graph\footnote{We also display an instance of the lollipop $L_{M,N}$ graph which is a fusion between a complete graph of size $M$ with a path of size $N$, and we run the SQWS on it in Appendix. \ref{app:sink}).} in Fig. \ref{fig:lollipop_graph}.

We first point out that for the complete, the hypercube and the cycle graphs, the search behavior will be the same no matter which vertex is marked, because these graphs are vertex-transitive, i.e. their structure does not allow vertices to be distinguished from each other. 

For the complete graph, we observe that all regimes $\omega$ achieve at least 50\% transfer efficiency to the sink when $\gamma\leq 2$. We see that for the purely quantum case, i.e. $\omega=0$, increasing the oracle strength with $\gamma$ improves transfer efficiency, reaching a maximum of 97\% for $\gamma= 1$. Then, when $\gamma>2$ is reached, transfer efficiency begins to decrease, reaching 0.02\% for $\gamma=30$. Interestingly, we see that even when $\gamma=0$ which corresponds to the free quantum walk with no oracle, the walker already achieves a transfer efficiency of 87\%. We also observe that the classical regime, i.e. $\omega=1$, achieves the highest transfer efficiency with 89\% for $\gamma=0$, which corresponds to a free quantum walk in the coherent regime, and does not induce a searching dynamics. Moreover, we note that variations in the parameter $\gamma$ do not seem to affect the hybrid regime when $\omega>0.1$, and furthermore, among the noisy regimes, we see that a higher value of $\omega$ results in better transfer to the sink. Thus, among the hybrid regimes, the worst transfer is obtained for $\omega=0.1$ and the quantum regime produces the worst transfer from $\gamma \geq 5$. For the hypercube, we obtain results similar to those for the complete graph, except that the maximum transfer efficiency achieved is lower, approximately 80\% for $\omega=0$. Furthermore, the performance of the purely quantum regime reaches its maximum of 84\% around $\gamma\approx 0.4$ and decreases sharply from $\gamma\approx 1$ to reach 0.01 for $\gamma=30$. For the hybrid regime, we observe the same phenomenon as for the complete graph, i.e. transfer efficiency increases with $\omega$, and variations in $\gamma$ only affect the hybrid regime from $\omega\geq 0.3$ onwards. For the other four graphs presented, we obtain different results. For the path graph we have run the SQWS on two different vertices that we call \textit{border} and \textit{center}, which are respectively placed at the extremity and center of the graph. For the maze, the target vertex corresponds to the exit and finally for the tadpole graph we chose three different vertices, \textit{cycle}, \textit{shared} and \textit{path}, their locations are shown on Fig. \ref{fig:lollipop_graph} on a smaller instance. First, the quantum regime is not the most efficient for any value of $\gamma$. Furthermore, the maximum transfer efficiency is low compared to the complete graph and the hypercube, as it does not exceed 50\%. The interesting phenomenon for these graphs is that, unlike in the previous cases, the regime around $\omega=0.1$ is the most efficient for all values of $\gamma$ tested. There is a clear and significant performance gain in this region when the dynamics consist of 90\% coherent evolution and 10\% non-unitary operations. For these graphs, the transfer to the sink is greatly improved by adding slight noise to the unitary dynamics.

In this study, we recover part of the results of Caruso et al. \cite{Caruso_2014,Caruso_2016} as their is an interesting phenomena around $\omega=0.1$ for specific graphs where the low-noise hybrid regime outperforms all the others. However, we only observe this phenomena for sparse graphs, and we obtain the opposite effect for the others, i.e. a significant drop in performance towards the $\omega=0.1$ regime compared to the other regimes. Therefore, the topology of the graph plays a major role in the performance of the SQWS and the parameter $\gamma$ mostly affect the quantum regime, i.e. $\omega=0$. In the quantum case, we observe a sharp decrease in performance when a critical value of $\gamma$ is reached. When this parameter continues to increase, we obtain a transfer of almost zero for the purely quantum regime, because the force of the oracle is too predominant over the exploration of the graph. Generally speaking, the SQWS behaves in a number of interesting ways, depending on the graph topology and the target vertex connectivity. The first behavior is the possibility for a low-noise hybrid regime to outperform quantum, classical and other hybrid dynamics for this modified search problem. We only observe this phenomenon for the cycle, path, maze and tadpole graphs. The second is the opposite behavior, where a low-noise hybrid regime is the worst for transfer efficiency in the context of low values of $\gamma$. This phenomenon is clearly observed for the complete graph and for the hypercube. Moreover, we see that for graphs where the quantum regime leads to a good transfer efficiency to the sink, when a critical value of $\gamma$ is reached, performance declines very rapidly, and thus a hybrid regime is more effective.

Intuitively, we can use metrics in an attempt to understand why the SQWS behaves differently on different graphs, and even on different vertices belonging to the same graph. A useful global graph metric is its density, with the densest complete graph serving as a reference with a density of 1. Then, two interesting local metrics for the target vertex are its eccentricity, i.e. the longest of the shortest paths to reach that vertex from any vertex in the graph, and its degree centrality, which indicates how connected the vertex is in the graph. A centrality of 1 means that the vertex is connected to all the others, and 0 to none. Interestingly, density and eccentricity are equal quantities for vertex transitive graphs, i.e. graphs whose structure does not allow vertices to be distinguished from each other. We show all these characteristics for all the graphs on which we have run the SQWS on Table \ref{tab:graphs}. The common feature of cycle, path, maze and tadpole graphs is that they all have very low density, and their vertices have high eccentricity and low degree centrality. These three characteristics appear to be necessary to observe the performance gain around $\omega=0.1$ that outperforms all other regimes. Furthermore, we note that for \textit{center} vertex of star and wheel graphs (see Appendix \ref{app:sink}), performance for $\omega=0.1$ does not decrease significantly as it does for other graphs, which may indicate that low density favours low noise regime performance around $\omega=0.1$. We also see that for the \textit{center} vertex of the wheel graph, performance is slightly lower for low values of $\gamma$ compared to the star graph, indicating that lower density and degree centrality favour low-noise regimes. Thus, the phenomenon observed by Caruso et al. \cite{Caruso_2014,Caruso_2016} seems to depend on the topology of the graph and is observed in this study for graphs with low density, poorly connected neighborhood, and which require long paths to reach a vertex from the target vertex.

\begin{table}
\caption{Global (density) and local metrics (degree centrality and eccentricity) of the target vertex for the different graphs on which we have run the Stochastic Quantum Walk Search (SQWS).}
\label{tab:graphs}
\centering
\renewcommand{\arraystretch}{1.5} 
\resizebox{\columnwidth}{!}{%
\begin{tabular}{c c c c c c}
\toprule
\textbf{Graph} & \textbf{Size} & \textbf{Density} & \textbf{Target vertex} $m$ & \textbf{Degree centrality} & \textbf{Eccentricity} \\ \toprule
Complete $K_{N}$ & $N=64$ & 1 & $\cdot$ & 1 & $1 = \mathcal{O}(1)$ \\ \hline
Cycle $C_{N}$ & $N=64$ & 0.0317 & $\cdot$ & 0.0317 & $32 = \lfloor N/2 \rfloor=\mathcal{O}(N)$ \\ \hline
\multirow{2}{*}{$d$-Hypercube $Q_{d}$} & \multirow{2}{*}{$N=2^{d}=64$} & \multirow{2}{*}{0.0952} & \multirow{2}{*}{$\cdot$} & \multirow{2}{*}{0.0952} & \multirow{2}{*}{$6 = d = \mathcal{O}(\log N)$} \\ 
               & ($d=6$) & & & & \\ \hline 
\multirow{2}{*}{Grid $G_{\sqrt{N}\times\sqrt{N}}$} & \multirow{2}{*}{$N=81$} & \multirow{2}{*}{0.0444} & \textit{center} & 0.05 & $8= \sqrt{N}-1 = \mathcal{O}(\sqrt{N})$ \\
               & & & \textit{border} & 0.025 & $16 = 2(\sqrt{N}-1)= \mathcal{O}(\sqrt{N})$ \\ \hline
\multirow{2}{*}{Star $S_{N-1}$} & \multirow{2}{*}{$N=64$} & \multirow{2}{*}{0.0312} & \textit{center} & 1 & $1 = \mathcal{O}(1)$ \\
               & & & \textit{border} & 0.0158 & $2 = \mathcal{O}(1)$ \\ \hline
\multirow{2}{*}{Wheel $W_{N}$} & \multirow{2}{*}{$N=64$} & \multirow{2}{*}{0.0625} & \textit{center} & 1 & $1 = \mathcal{O}(1)$ \\
               & & & \textit{border} & 0.0476 & $2 = \mathcal{O}(1)$ \\ \hline
\multirow{3}{*}{Perfect Binary Tree} & \multirow{3}{*}{$N\!=\!2^{d+1}\!-\!1\!=\!63$} & \multirow{3}{*}{0.0317} & $d_m = 0$ (\textit{root}) & 0.0322 & $5 = d + d_m = \mathcal{O}(\log N)$ \\
               & & & $d_m = 3$ & 0.0483 & $8 =d+d_m= \mathcal{O}(\log N)$ \\ 
               $PBT_{d}$ of depth $d$ & ($d=5$) & & $d_m = 5$ (\textit{leaf}) & 0.0161 & $10 =d+d_m= \mathcal{O}(\log N)$ \\ \hline
\multirow{2}{*}{Path $P_{N}$} & \multirow{2}{*}{$N=65$} & \multirow{2}{*}{0.0307} & \textit{center} & 0.0312 & $32 =\lfloor N/2\rfloor= \mathcal{O}(N)$ \\ 
               & & & \textit{border} & 0.0156 & $64 =N-1= \mathcal{O}(N)$ \\ \hline
\multirow{3}{*}{Lollipop $L_{M,N}$} & \multirow{3}{*}{$M$+$N\!=\!32$+$32$} & \multirow{3}{*}{0.2619} & \textit{complete} & 0.4920 & $33 = N+1=\mathcal{O}(N)$ \\ 
               & & & \textit{shared} & 0.5079 & $32 = N=\mathcal{O}(N)$ \\ 
               & & & \textit{path} & 0.0158 & $33 = N+1=\mathcal{O}(N)$ \\ \hline
\multirow{3}{*}{Tadpole $T_{M,N}$} & \multirow{3}{*}{$M$+$N\!=\!32$+$32$} & \multirow{3}{*}{0.0317} & \textit{cycle} & 0.0317 & $48\!=\!N+\lfloor M/2\rfloor= \mathcal{O}(N$+$M)$ \\ 
               & & & \textit{shared} & 0.0476 & $32\!=\!\max(N,\lfloor M/2\rfloor)\!=\!\mathcal{O}(N$+$M)$\! \\ 
                 & & & \textit{path} & 0.0158 & $48\!=\!N+\lfloor M/2\rfloor\!=\!\mathcal{O}(N$+$M)$ \\ \hline
\multirow{3}{*}{Random (Small-World)} & \multirow{3}{*}{$N=66$} & \multirow{3}{*}{0.0867} & \textit{HC} & 0.1846 & 6 \\ 
               & & & \textit{IC} & 0.0923 & 6 \\
             $SW_N$  & & & \textit{LC} & 0.0615 & 5 \\ \hline
Maze $M_{N}$ & $N=73$ & 0.0273 & \textit{exit} &  0.0138 & 34 \\ \bottomrule
\end{tabular}
}
\end{table}

\begin{figure}
    \centering
    \includegraphics[width=0.4\textwidth]{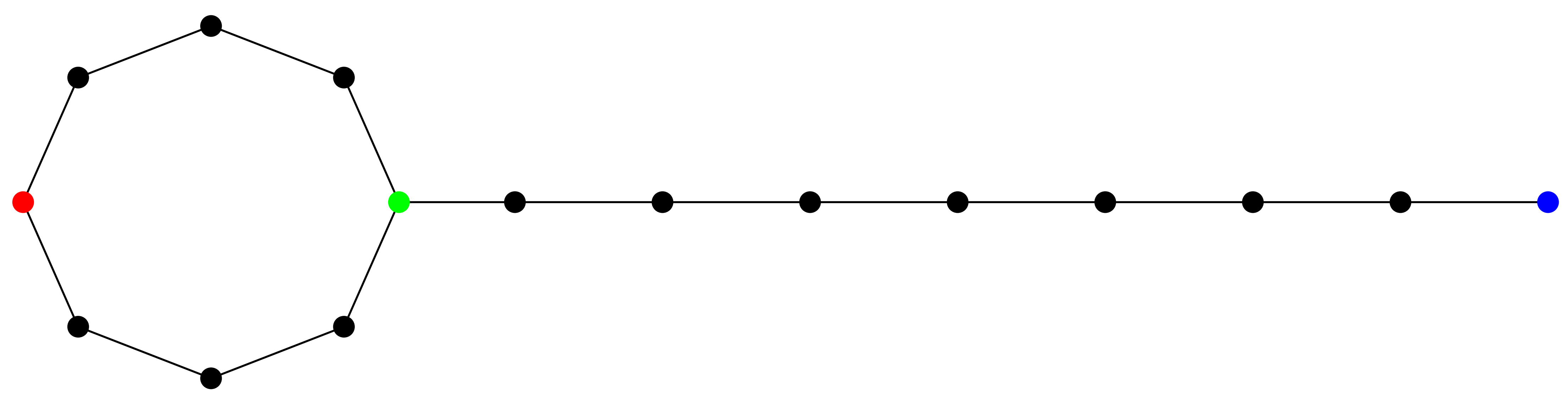}
    \includegraphics[width=0.4\textwidth]{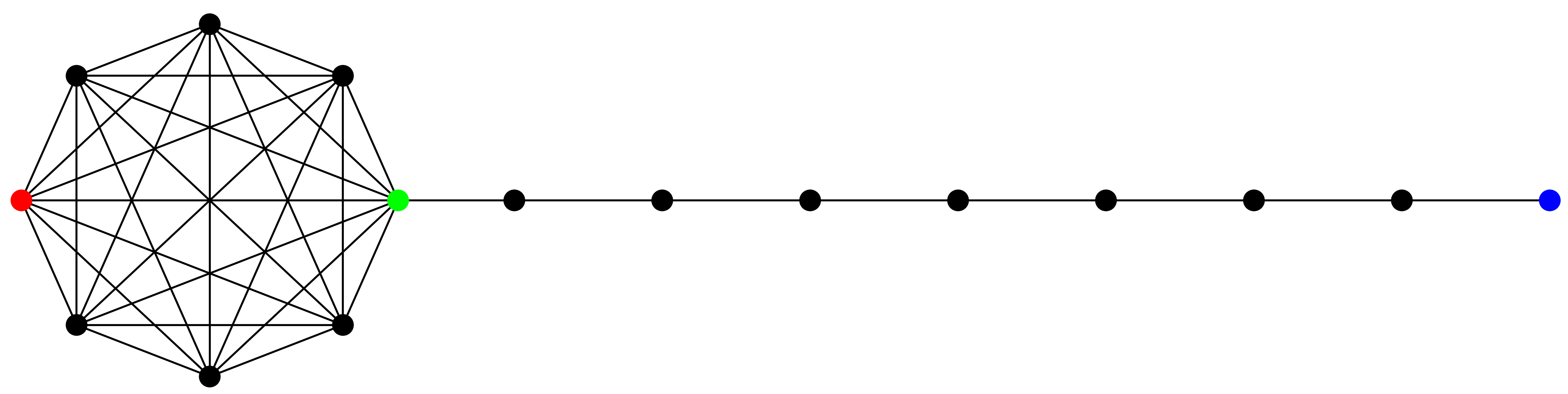}
    \caption{Instances $T_{8,8}$ and $L_{8,8}$ of the tadpole (upper) and lollipop (lower) graphs, that are respectively a fusion between the cycle graph $C_8$ or the complete graph $K_8$ with the path graph $P_8$. We respectively refer to the locations of the red, green and blue vertices as \textit{cycle}, \textit{shared} and \textit{path} vertices for the tadpole graph, and \textit{complete}, \textit{shared} and \textit{path} for the lollipop graph.}
    \label{fig:lollipop_graph}
\end{figure}

We explore the behavior of the SQWS on many different families of graphs, and observe how it behaves when a cycle is gradually transformed into a complete graph in Appendix \ref{app:sink}.

\subsection{Relation to entropy}

\begin{figure}
\centering
\begin{subfigure}{0.46\textwidth}
    \includegraphics[width=\textwidth]{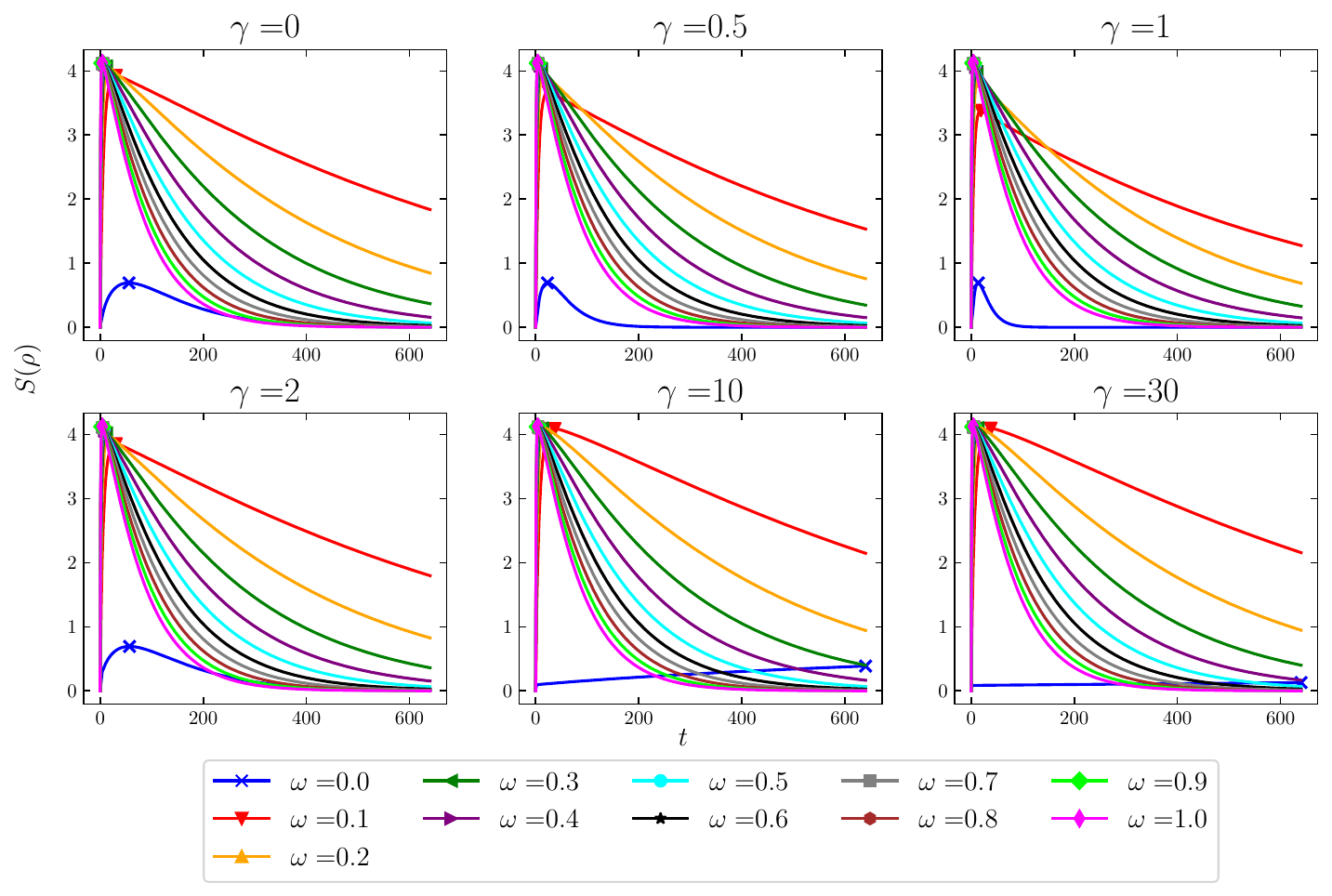}
    \caption{Complete graph}
    \label{fig:entropy_complete}
\end{subfigure}
\begin{subfigure}{0.46\textwidth}
    \includegraphics[width=\textwidth]{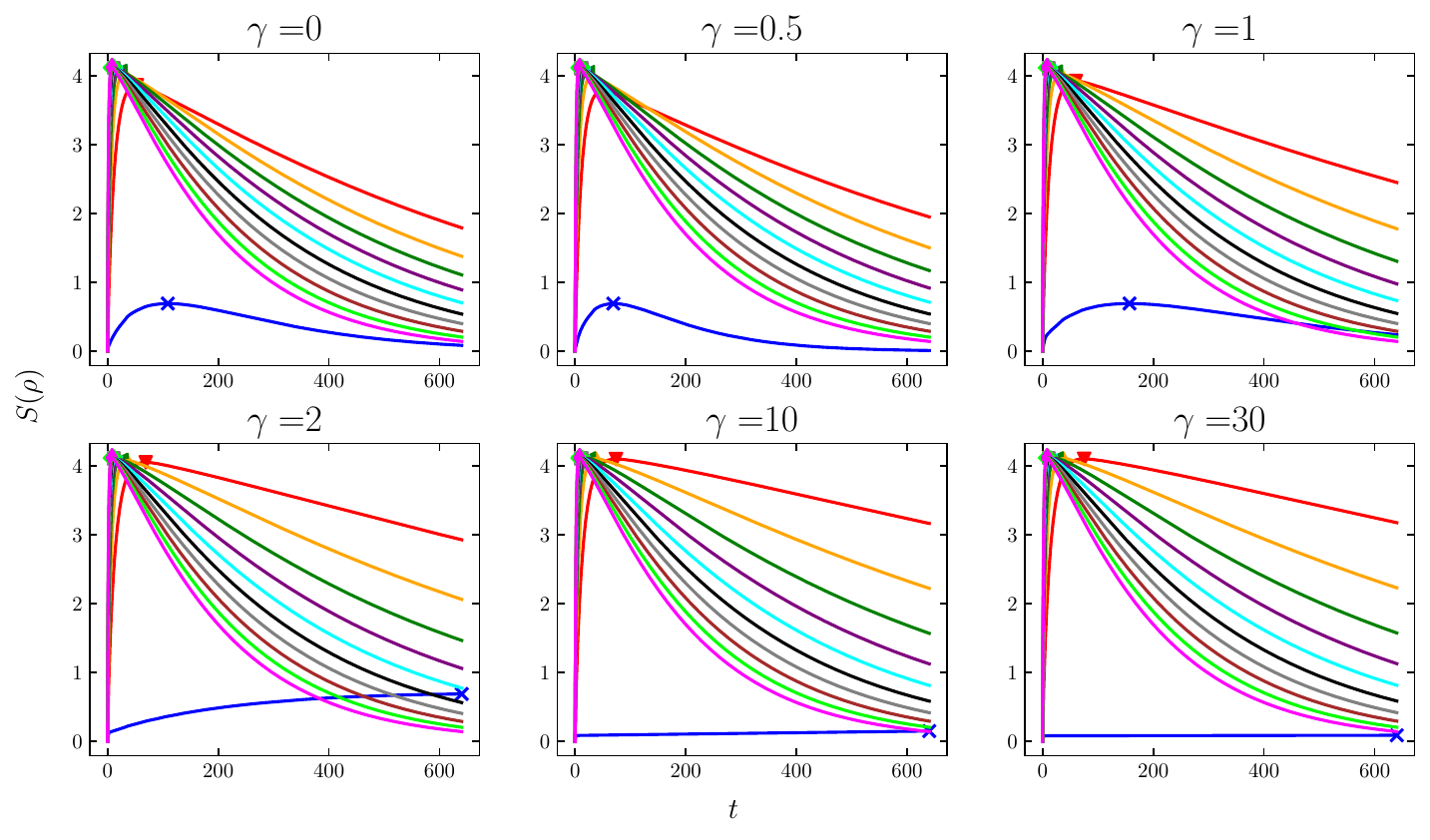}
    \caption{Hypercube}
    \label{fig:entropy_hypercube}
\end{subfigure}
\begin{subfigure}{0.46\textwidth}
    \includegraphics[width=\textwidth]{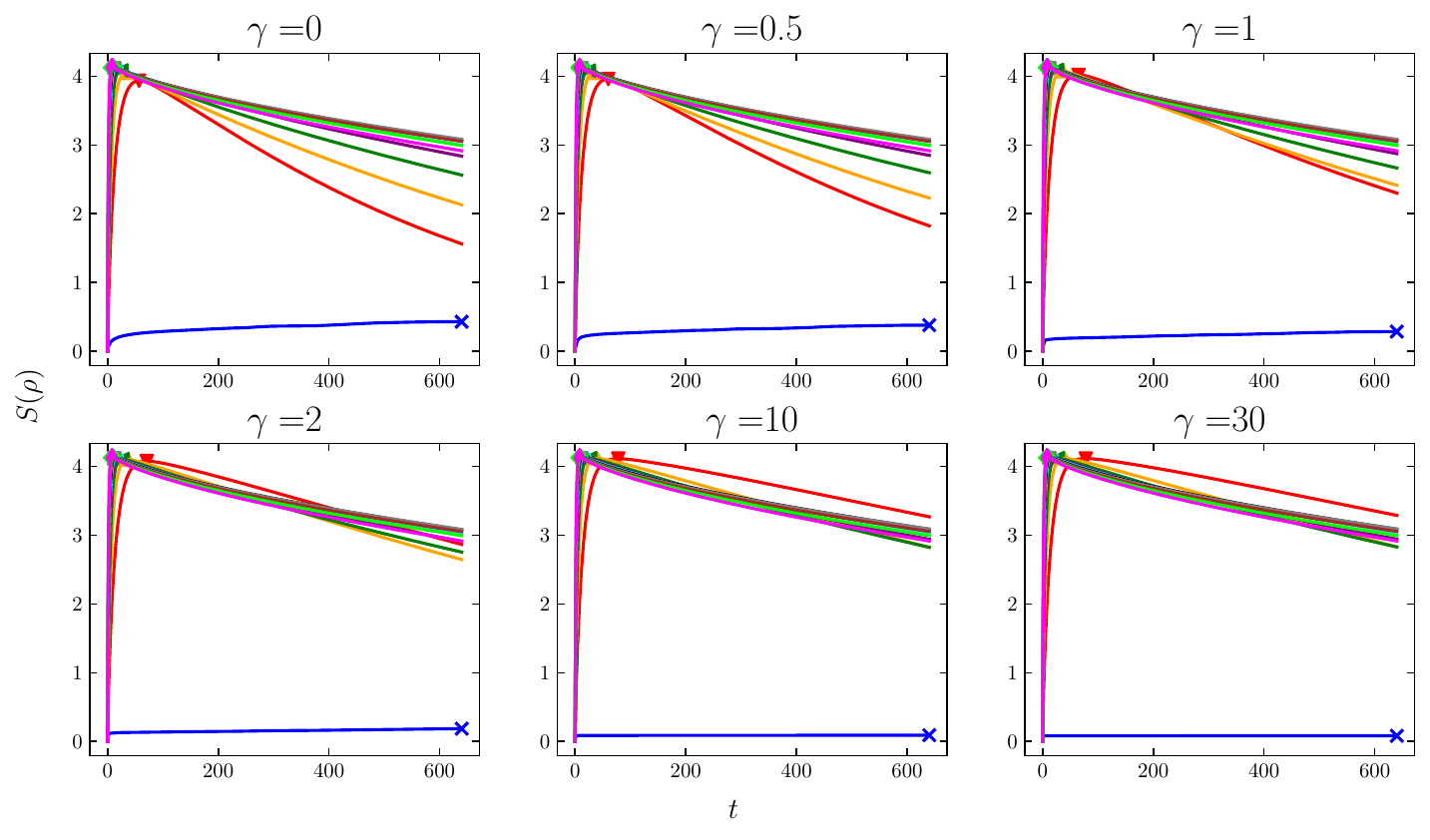}
    \caption{Cycle graph}
    \label{fig:entropy_cycle}
\end{subfigure}
\begin{subfigure}{0.46\textwidth}
    \includegraphics[width=\textwidth]{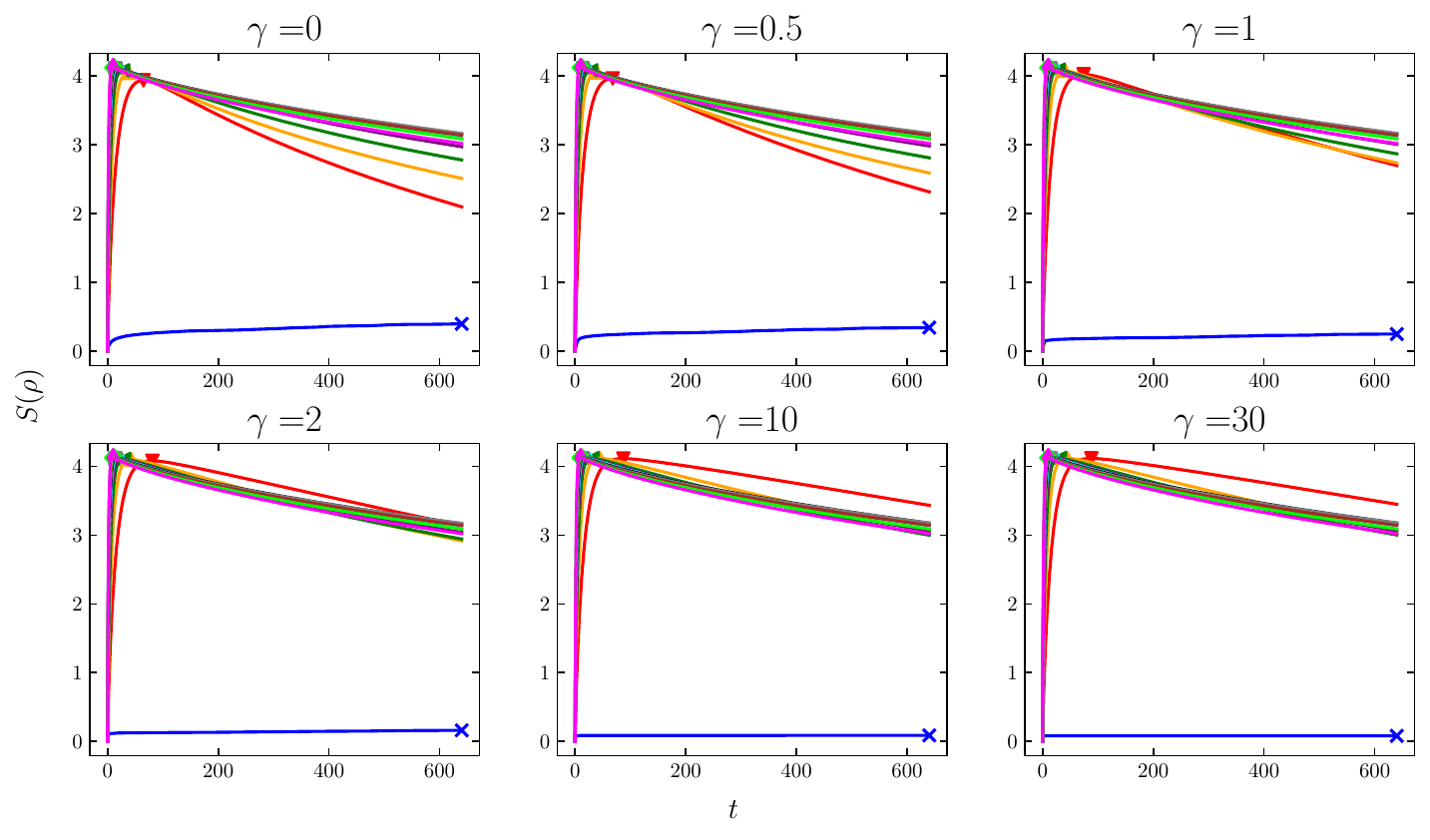}
    \caption{Tadpole graph \textit{cycle} vertex}
    \label{fig:entropy_tadpole}
\end{subfigure}
\caption{Von Neumann entropy $S(\rho)$ as a function of time $t$ for different values of $\gamma$ for the complete graph $K_{64}$, the 6-hypercube $Q_6$, the cycle graph $C_{64}$ and the tadpole graph $T_{32,32}$ for the \textit{cycle} vertex. The markers indicate the time $t_S$ at which the entropy reaches its maximum value before decreasing down to zero.}
\label{fig:entropy}
\end{figure}

We now relate the performance of an interpolation regime $\omega$ to the evolution of the Von Neumann entropy of the system:
\begin{equation}
    S(\rho) = -\tr{\rho\ln\rho}.
\end{equation}
We observe for every run of the SQWS that the entropy first increases up to a maximum, and then decreases until it converges to zero. We also observe that when an increase in the value of $\gamma$ increases the performance of the SQWS, i.e. the transfer to the sink, this translates into a reduction in the time needed to reach the entropy maximum and a faster convergence to zero thereafter. Furthermore, we observe that the interpolation regime $\omega$ whose entropy converges to zero the fastest has the highest transfer efficiency. The presence of the sink vertex $\phi$ introduces dissipation in the system, therefore, as $t\rightarrow\infty$ the walker will end up in the sink with a probability of 1. Thus, the state of the walker converges from $\rho(0)$ to the projector $\ket{\phi}\bra{\phi}$. Although zero entropy indicates a pure state, it does not guarantee that this state is the state towards which the system converges. Therefore, we also compute the $l_1$-norm coherence, which is the sum of the off-diagonal elements of $\rho$:
\begin{equation}
    C_{l_1}(\rho) = \sum_{i\neq j}|\rho_{ij}|.
\end{equation}
As the initial state is the uniform superposition over the set of vertices $V$, the initial value is $C_{l_1}(\rho(0))=|V|-1$. For all graphs on which we run the SQWS, we observe that this quantity decreases monotonically and it converges to zero. Thus, the convergence of entropy to zero does indicate that the walker state is getting closer and closer to the state $\ket{\phi}\bra{\phi}$, as $\rho$ becomes increasingly diagonal over time. We illustrate the entropy evolution over time in Fig. \ref{fig:entropy} for the complete, the hypercube, the cycle and the tadpole (for the \textit{cycle} vertex) graphs. 

Looking at Fig. \ref{fig:entropy_complete}, we see that for the complete graph the regime with entropy converging most rapidly to zero for $\gamma=0$ is $\omega=1$. Indeed, the random walk search achieves a transfer efficiency of 89\% compared to 87\% for $\gamma=0$. We then see that from $\gamma=0.5$ onwards, the quantum regime converges more quickly, as it achieves an efficiency of 94\%, then converges even more quickly for $\gamma=1$, where it reaches its maximum efficiency of 97\%. We also observe that from $\gamma=2$ onwards, convergence towards zero becomes slower, then when $\gamma$ continues to increase, the chosen evolution time is not sufficient for the entropy to reach its maximum and begin to converge towards zero, which results in very poor performance. It is also clear that the $\omega=0.1$ regime performs less well than the others due to its slow convergence. For the hypercube, we see on Fig. \ref{fig:entropy_hypercube} that the quantum regime is the most efficient for $\gamma=0$, and is already less efficient than the classical one for $\gamma=1$ as it can be seen on Fig. \ref{fig:qsws}. From $\gamma=2$ onwards, we see that the entropy of the quantum regime no longer reaches its maximum with the chosen evolution time, and the most efficient regimes are those with the highest values of $\omega$. As expected, we observe different behaviours for the cycle graph and the \textit{cycle} vertex of the tadpole graph \footnote{Note that the same results occur for the maze, the path and the other vertices of the tadpole, but we only display two of them.} on Figs. \ref{fig:entropy_cycle} and \ref{fig:entropy_tadpole}. This time, we see that the weakly noisy regime is the most effective for low values of $\gamma$, while the other regimes, except for the quantum regime, see their entropy decrease more rapidly with increasing $\gamma$ and eventually decrease more rapidly than $\omega=0.1$. Finally, we observe that entropy converges more slowly than for the complete graph and the hypercube, which explains the lower maximum transfer efficiency levels achieved.

In general, we can clearly see from Fig. \ref{fig:entropy} that the efficiency of transfer to the sink translates firstly into a reduction in the time $t_S$ needed to reach maximum entropy. Then, by a convergence of entropy towards zero as quickly as possible, where the $\omega$-regime maximizing efficiency is the one with the fastest convergence.

\begin{figure}
    \centering
    \includegraphics[width=0.5\textwidth]{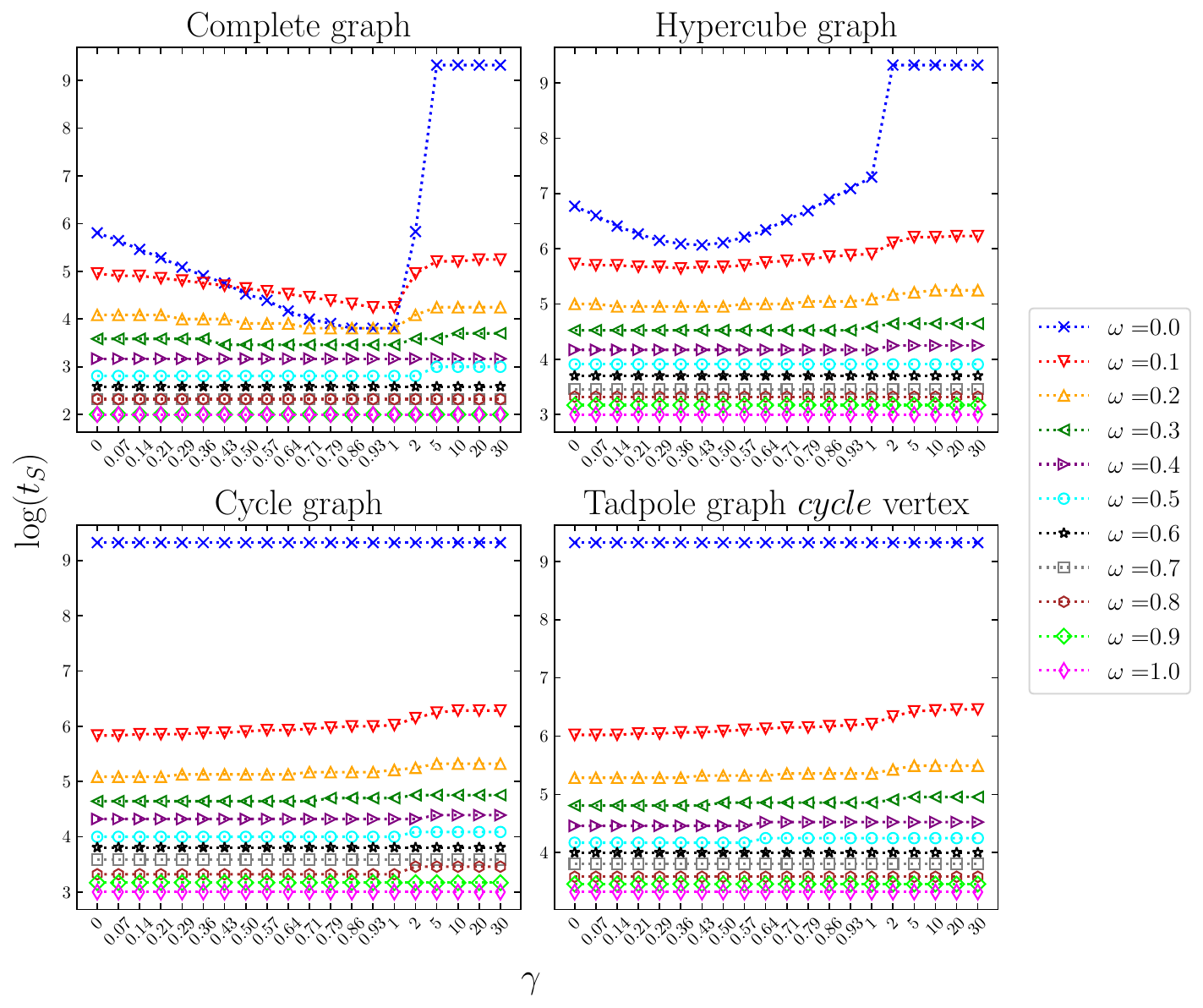}
    \caption{Evolution of the duration $t_S$ needed to reach maximum entropy as a function of $\gamma$ for the complete graph $K_{64}$, the 6-hypercube $Q_6$, the cycle graph $C_{64}$ and the tadpole graph $T_{32,32}$ (with \textit{cycle} vertex) for different values of interpolation $\omega\in[0,1]$.}
    \label{fig:time_entropy}
\end{figure}

We also illustrate the reduction of duration $t_S$ required to reach the maximum entropy in Fig. \ref{fig:time_entropy} for these graphs. We can see that for the complete graph and the hypercube, variations in $\gamma$ mainly affect the duration $t_S$ for $\omega=0$ and $\omega=0.1$. The most sensitive regime is the quantum regime, and we observe that the duration $t_S$ reaches its minimum for $\gamma=1$ and $\gamma=0.43$, respectively, which correspond to the values of $\gamma$ that give the highest transfer efficiency for these graphs. We also see that once these values are exceeded, this duration increases dramatically, which delays the moment when entropy will decrease to begin converging towards zero and thus results in a significant loss of performance. For the cycle and tadpole graphs, variations in $\gamma$ have no effect on the quantum regime and seem to affect only the low-noise regime for $\omega=0.1$ and $\omega=0.2$.

\section{Conclusion}

In summary, we have studied a continuous-time search problem in which the walker explores a graph according to a searching dynamics driven to a specific target vertex. We also connected a trapping sink vertex to the target with an irreversible transition. We were interested in the transfer efficiency of a slightly modified search problem, and not in the its computational complexity of the search problem. We extended the result of Caruso et al. \cite{Caruso_2014,Caruso_2016} to a Stochastic Quantum Walk Search (SQWS) and we numerically showed that a tunable mixing of unitary and non-unitary dynamics can lead to a better transfer efficiency than a non-hybrid evolution for this problem depending on (i) the graph topology, (ii) the target vertex connectivity and (iii) a parametrized Hamiltonian. In particular, the Hamiltonian parameter controls the strength of the oracle that marks the target vertex. We have also related performance of an interpolation regime to the system entropy decay rate. Moreover, we have shown that the hybrid regime can beat the purely quantum dynamics only in the presence of a trapping sink. 
For numerous graphs, mostly dense, quantum evolution is very sensitive to an increase in the strength of the oracle which is not the case for the hybrid regime. We have also shown that low-noise regimes give the worst performance for most of the graphs, expect for sparse graphs, with low degree centrality and high eccentricity for the target vertex, where these regimes are the most efficient when the strength of the oracle remains fairly weak. Therefore, by considering the value of this parameter as a computational resource, the hybrid evolution may require fewer resources than the quantum to perform on specific topologies. More importantly, at a fixed parameters configuration, we can still play with the interaction graph to fit the optimal transfer performance. 

This can pave the way to a technological leap where noise can be seen as a useful physical resource, in a hardware setting where one can change the connectivity of the architecture for reliable quantum computing. In conclusion, future work could also focus to provide a natively circuit based model for the above results, introducing quantum noise as close as possible to real physical devices'.

\section{Data availability}

Simulations were carried out with the Python library QuTip~\cite{Johansson_2012} and the graphs were generated with NetworkX~\cite{hagberg2008exploring}. The code is available at: \href{https://github.com/ugo-nzongani/Stochastic-Quantum-Walk-Search}{SQWS}.

\section{Acknowledgements}

We thank Ravi Kunjwal, Joachim Tomasi, Julien Zylberman and Sunheang Ty for their usefull feedback on the form and content of this manuscript. This work is supported by the PEPR EPiQ ANR-22-PETQ-0007, by the ANR JCJC DisQC ANR-22-CE47-0002-01.

\appendix

\section{SQWS with no sink}\label{app:no_sink}

\begin{figure*}
    \centering
    \includegraphics[width=0.8\textwidth]{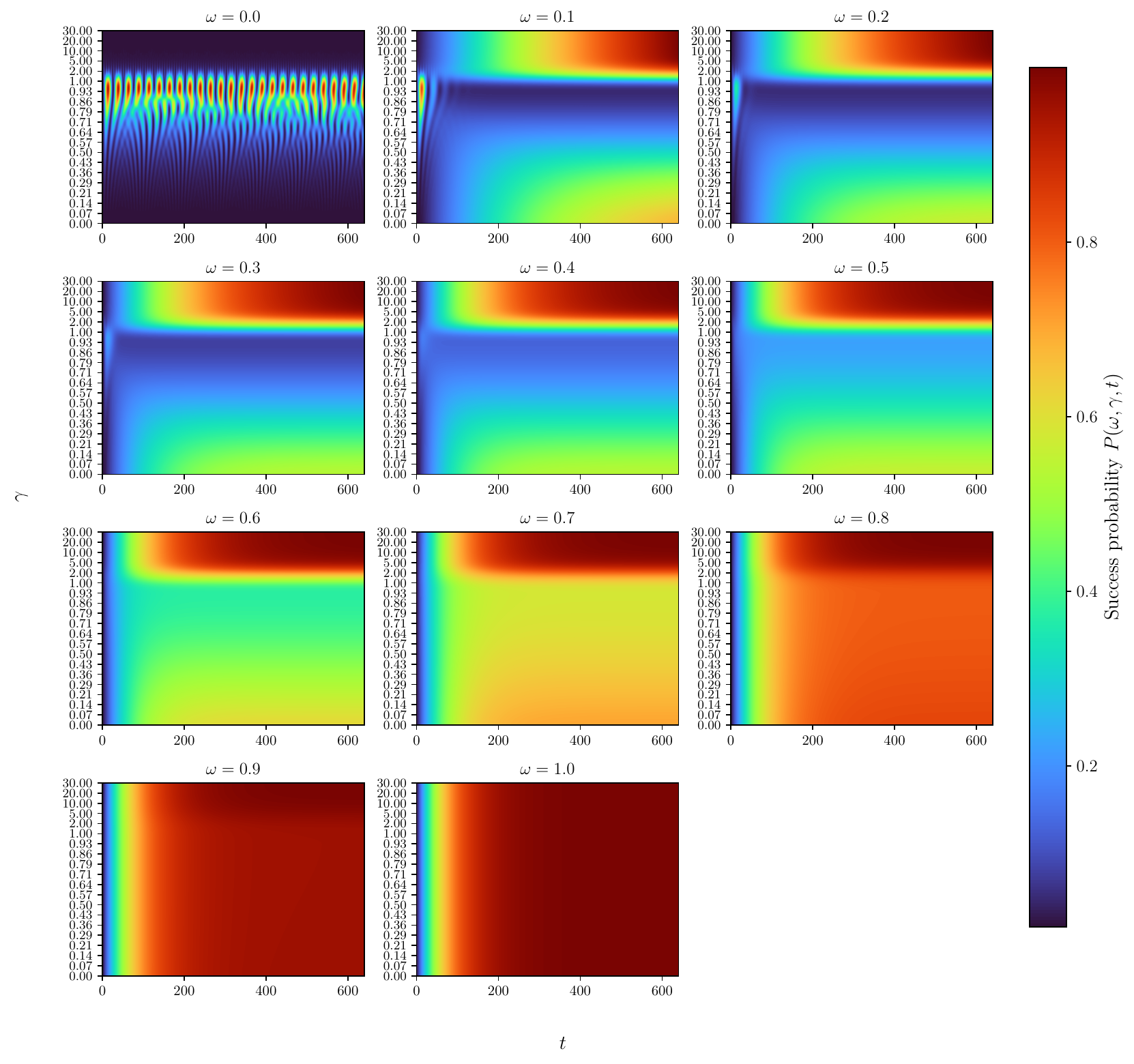}
    \caption{Probability $P(\omega,\gamma,t)$ over time of finding the target vertex using the Stochastic Quantum Walk Search (SQWS) with no sink vertex connected to the target vertex, i.e. $\Gamma=0$ in Eq. \eqref{eq:qsw_search}, on the complete graph ($K_{64}$). The quantum walk search is recovered for $\omega=0$, the classical random walk search for $\omega=1$, and a linear combination of the two when $\omega\in]0,1[$.}
    \label{fig:no_sink_complete}
\end{figure*}

\begin{figure*}
    \centering
    \includegraphics[width=0.8\textwidth]{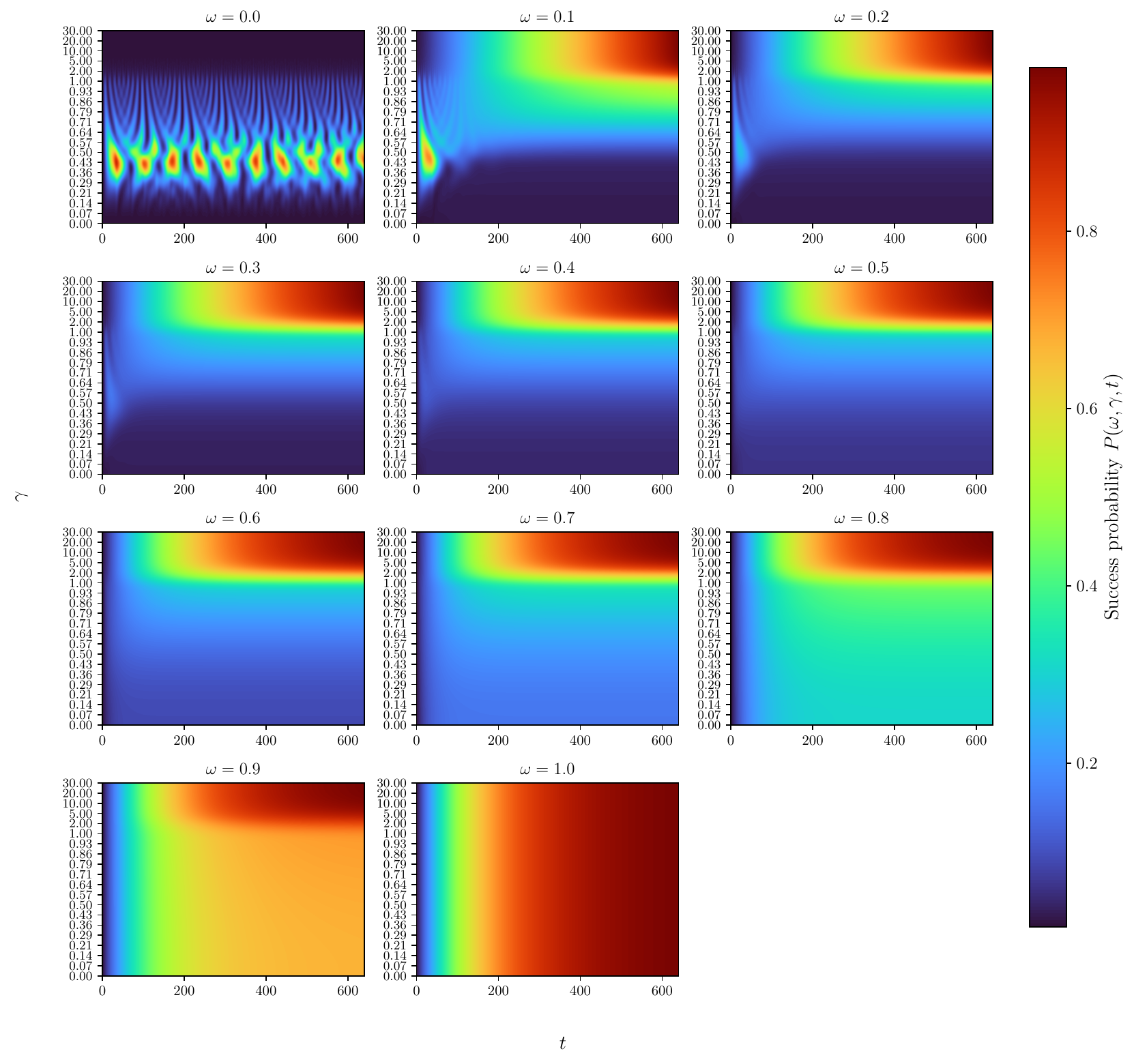}
    \caption{Probability $P(\omega,\gamma,t)$ over time of finding the target vertex using the Stochastic Quantum Walk Search (SQWS) with no sink vertex connected to the target vertex, i.e. $\Gamma=0$ in Eq. \eqref{eq:qsw_search}, on the hypercube graph ($Q_{6}$). The quantum walk search is recovered for $\omega=0$, the classical random walk search for $\omega=1$, and a linear combination of the two when $\omega\in]0,1[$.}
    \label{fig:no_sink_hypercube}
\end{figure*}

\begin{figure*}
    \centering
    \includegraphics[width=0.8\textwidth]{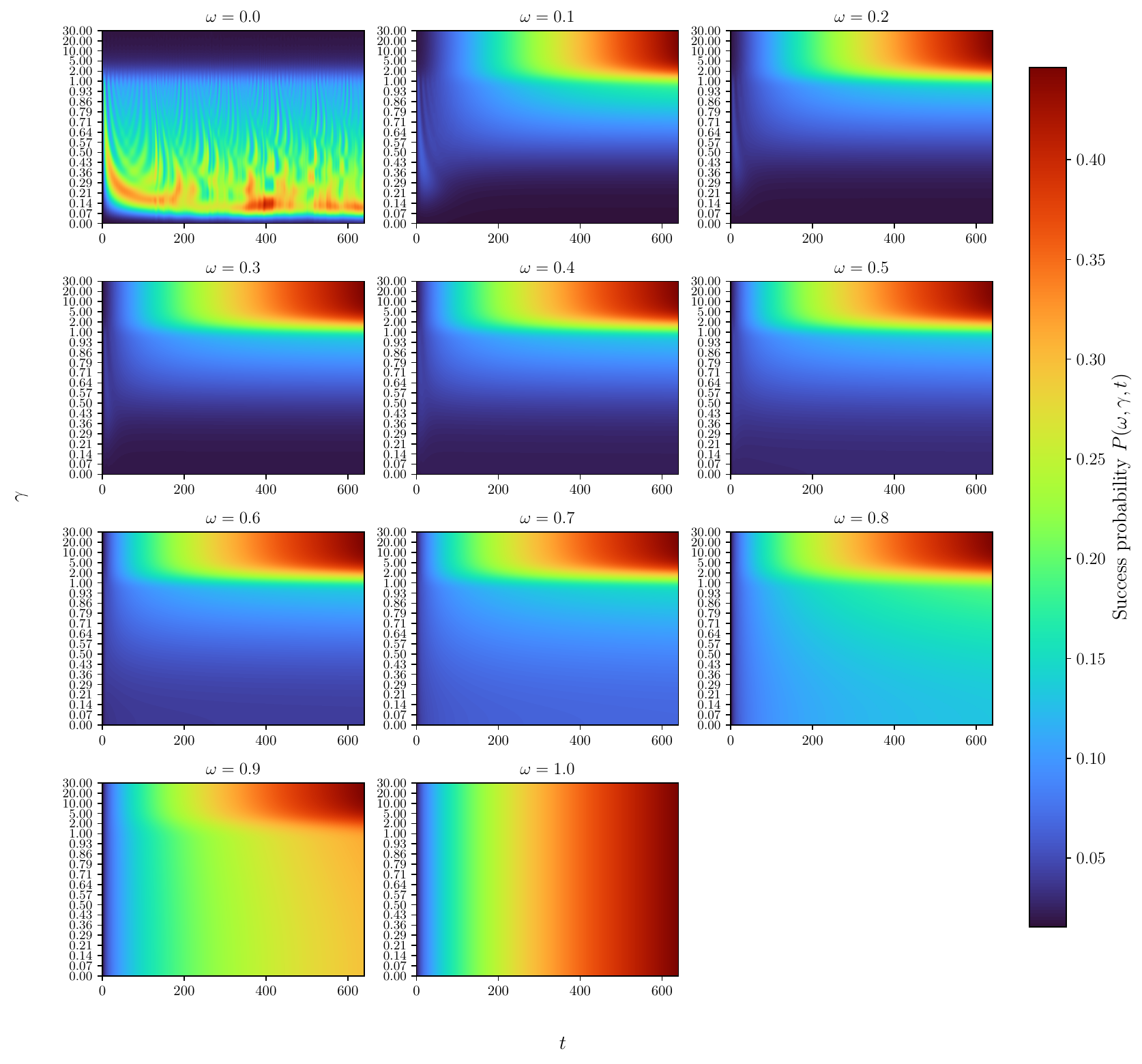}
    \caption{Probability $P(\omega,\gamma,t)$ over time of finding the target vertex using the Stochastic Quantum Walk Search (SQWS) with no sink vertex connected to the target vertex, i.e. $\Gamma=0$ in Eq. \eqref{eq:qsw_search}, on the cycle graph ($C_{64}$). The quantum walk search is recovered for $\omega=0$, the classical random walk search for $\omega=1$, and a linear combination of the two when $\omega\in]0,1[$.}
    \label{fig:no_sink_cycle}
\end{figure*}

In this section, we briefly discuss the performances of the SQWS with no use of an extra sink vertex connected to the target vertex. We set $\Gamma=0$ in Eq. \eqref{eq:qsw_search}. The success probability is now related to the presence of the walker in the target vertex $m$ instead of the sink:
\begin{equation}
    P(\omega,\gamma,t) = \tr{\mathcal{E}[\rho(t)]\ket{m}\bra{m}}.
\end{equation}
Childs and Goldstone search provides a quadratic speedup for the complete graph and the hypercube~\cite{childs2004spatial}, meaning that the evolved state $\mathcal{E}[\rho(t)]$ reaches a large overlap with the target $\ket{m}\bra{m}$ in a timeframe that scales as $t=\mathcal{O}(\sqrt{|V|})$, which is not the case for the cycle graph~\cite{chakraborty2020optimality}. Therefore, we run the SQWS on instances of the complete graph, the hypercube and the cycle graph and respectively show the results in Figs. \ref{fig:no_sink_complete}, \ref{fig:no_sink_hypercube} and \ref{fig:no_sink_cycle}. For the complete graph, we find an oscillatory behaviour for the probability of success when $\omega=0$ because the dynamics are completely unitary. We see that the highest performance is obtained for $\gamma\in[0.8,1.5].$ Then, as soon as $\omega$ increases, we lose the oscillatory behaviour and the quantum regime no longer performs well. Furthermore, for the hybrid regime, we see that we obtain a probability of success of 0.8 for much longer evolution times and higher values of $\gamma$ than were necessary to obtain the same probability with the quantum regime. When $\omega= 0.8$, we obtain a probability of success of approximately 0.7 for values of $\gamma<1$. Finally, we obtain almost similar results for $\omega=0.9$ and $\omega=1$, although the time required to reach a 0.8 probability of success is much higher than that required for the quantum regime. Similar results are obtained for the hypercube, except that the optimal values of $\gamma$ are lower than for the complete graph for the quantum regime. Otherwise, for the hybrid regime, we observe more or less the same results, and we must wait for the classical regime $\omega=1$ to achieve a probability of success of approximately 0.8 for values of $\gamma<1$, although the evolution time required is much longer than for the quantum regime. For the graph cycle, we obtain different results for the quantum regime. The highest probability of success is 0.4 and is achieved for a considerably higher time value than for the two previous graphs. For the hybrid and classical regimes, we obtain results similar to the previous graphs, although an even longer evolution time is required to reach the maximum probability of success of approximately 0.4.

These results indicate that the presence of the sink is necessary to observe phenomena where the hybrid case outperforms the quantum regime, because without its presence, the addition of non-unitary dynamics only degrades performance.

\section{SQWS with sink}\label{app:sink}

In this section, we take the numerical study of the SQWS a step further by running it on numerous instances of different graph families.

\subsection{Additional graphs}

\begin{figure*}
    \centering
    \includegraphics[width=0.3\textwidth]{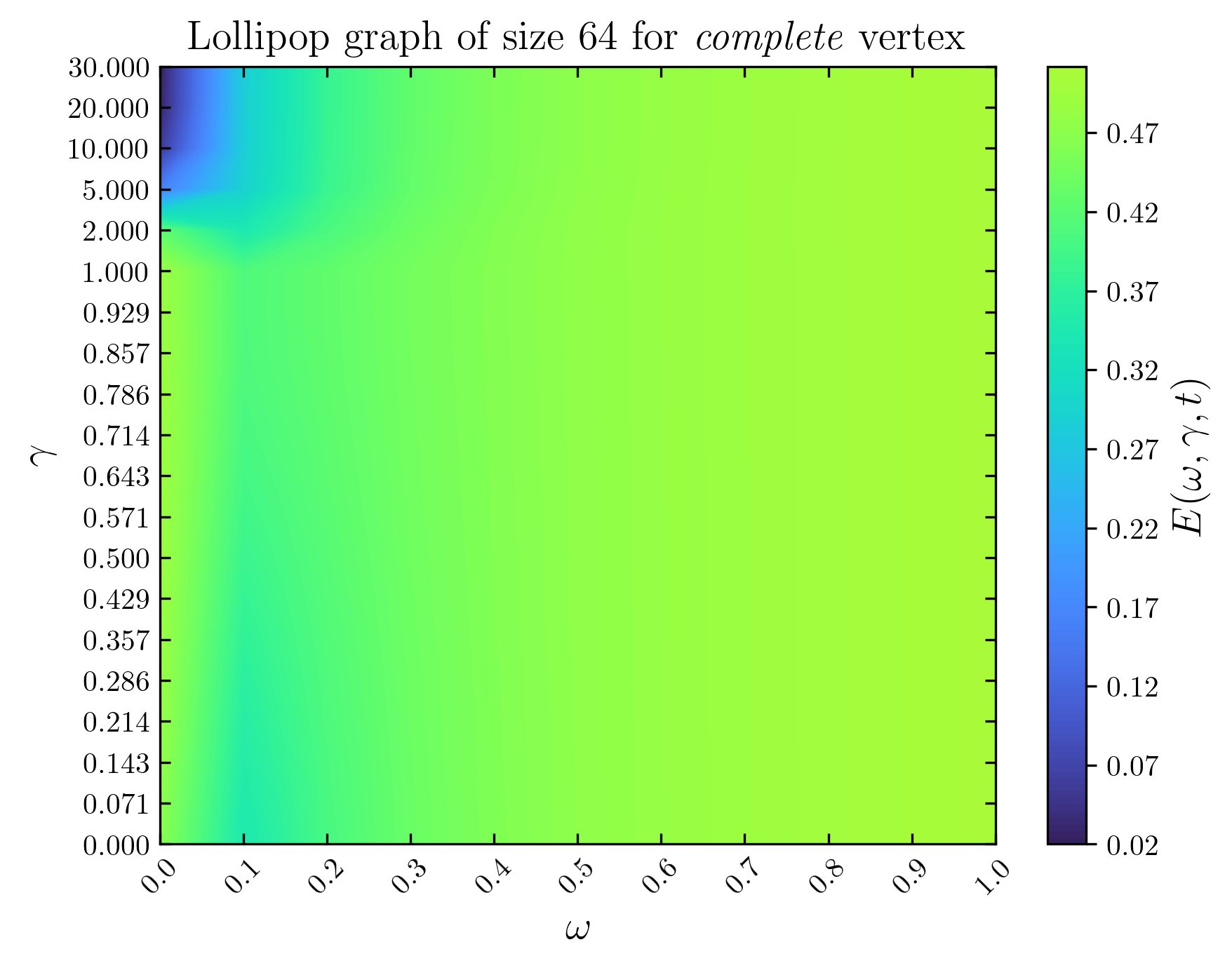}
    \includegraphics[width=0.3\textwidth]{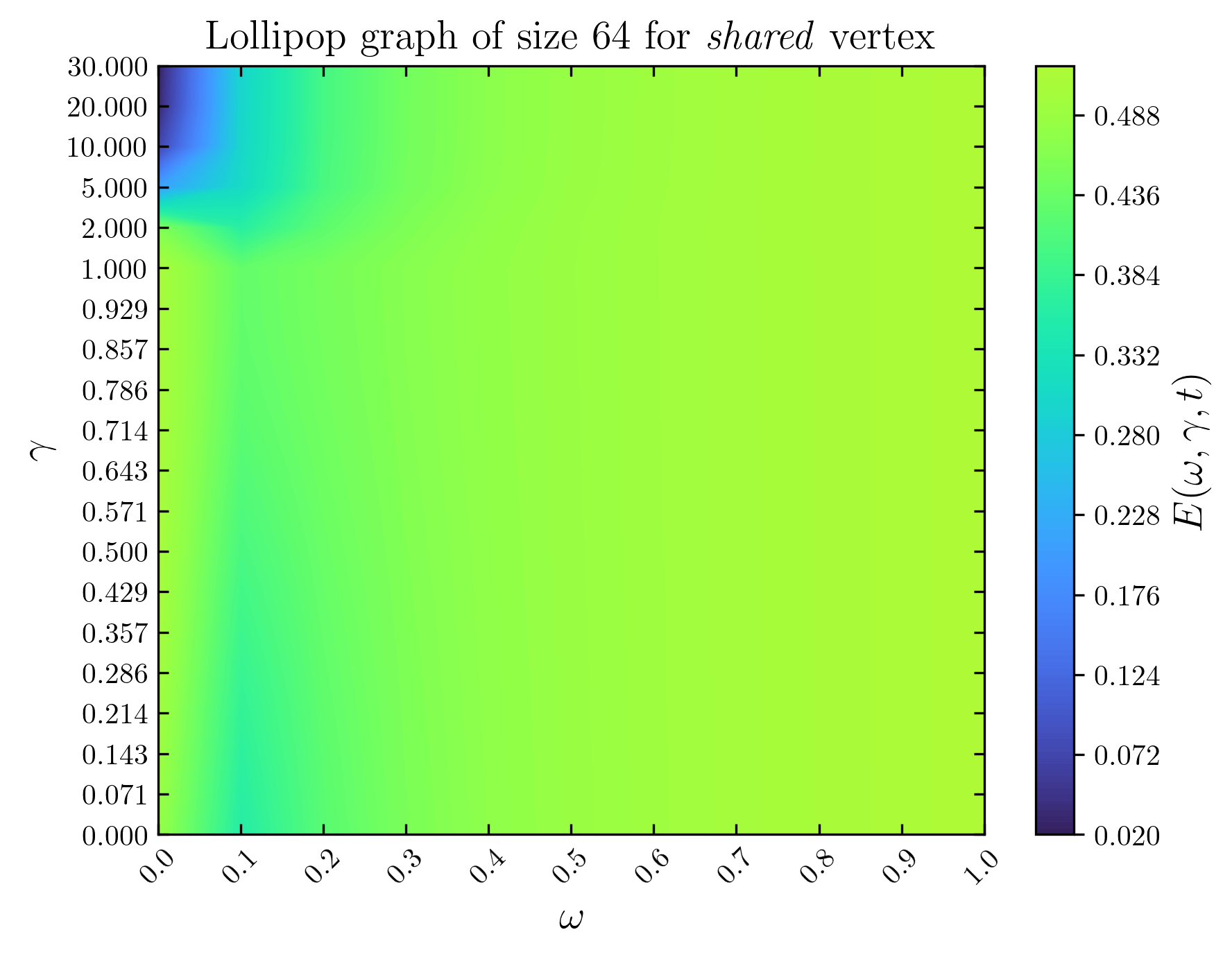}
    \includegraphics[width=0.3\textwidth]{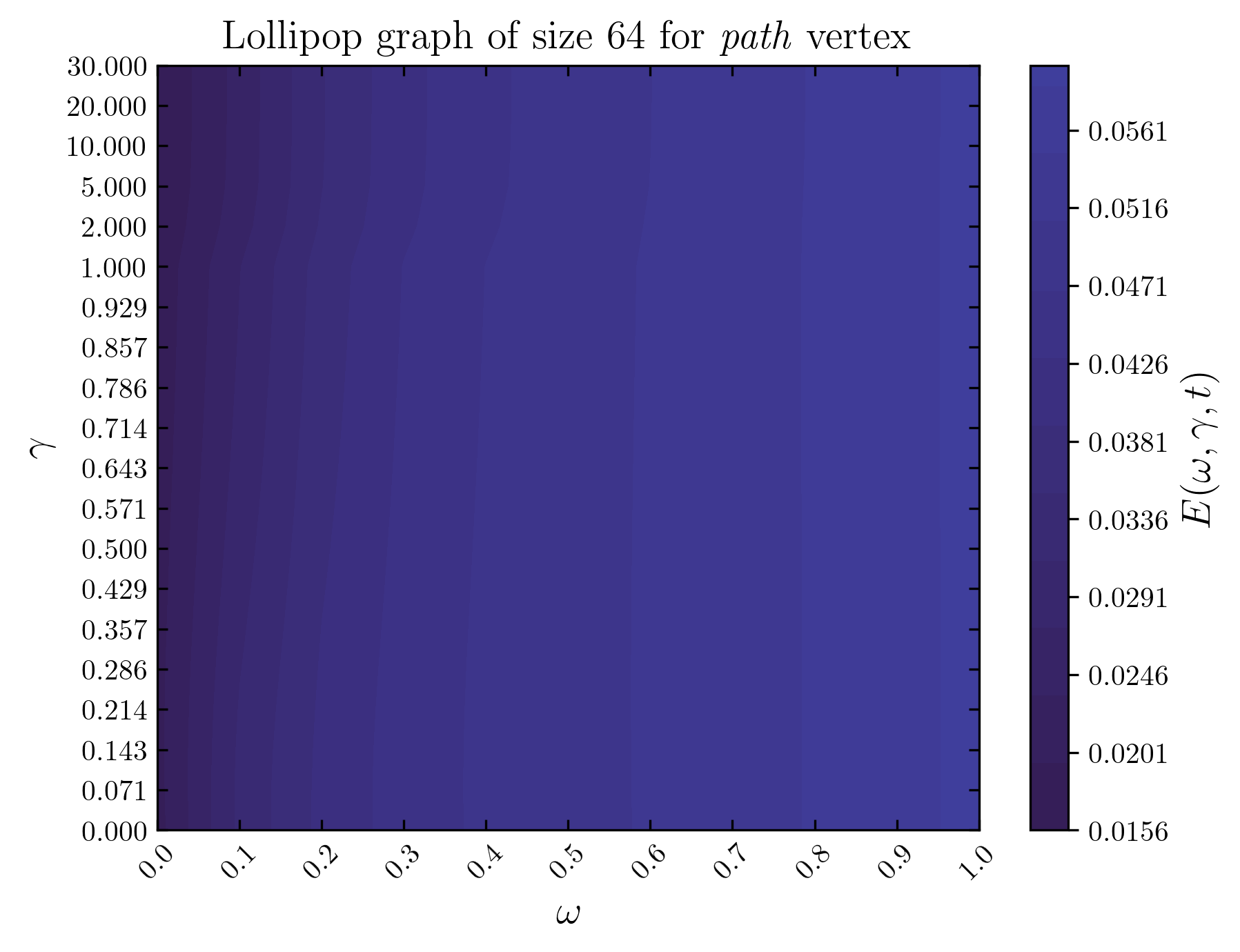}
    \includegraphics[width=0.3\textwidth]{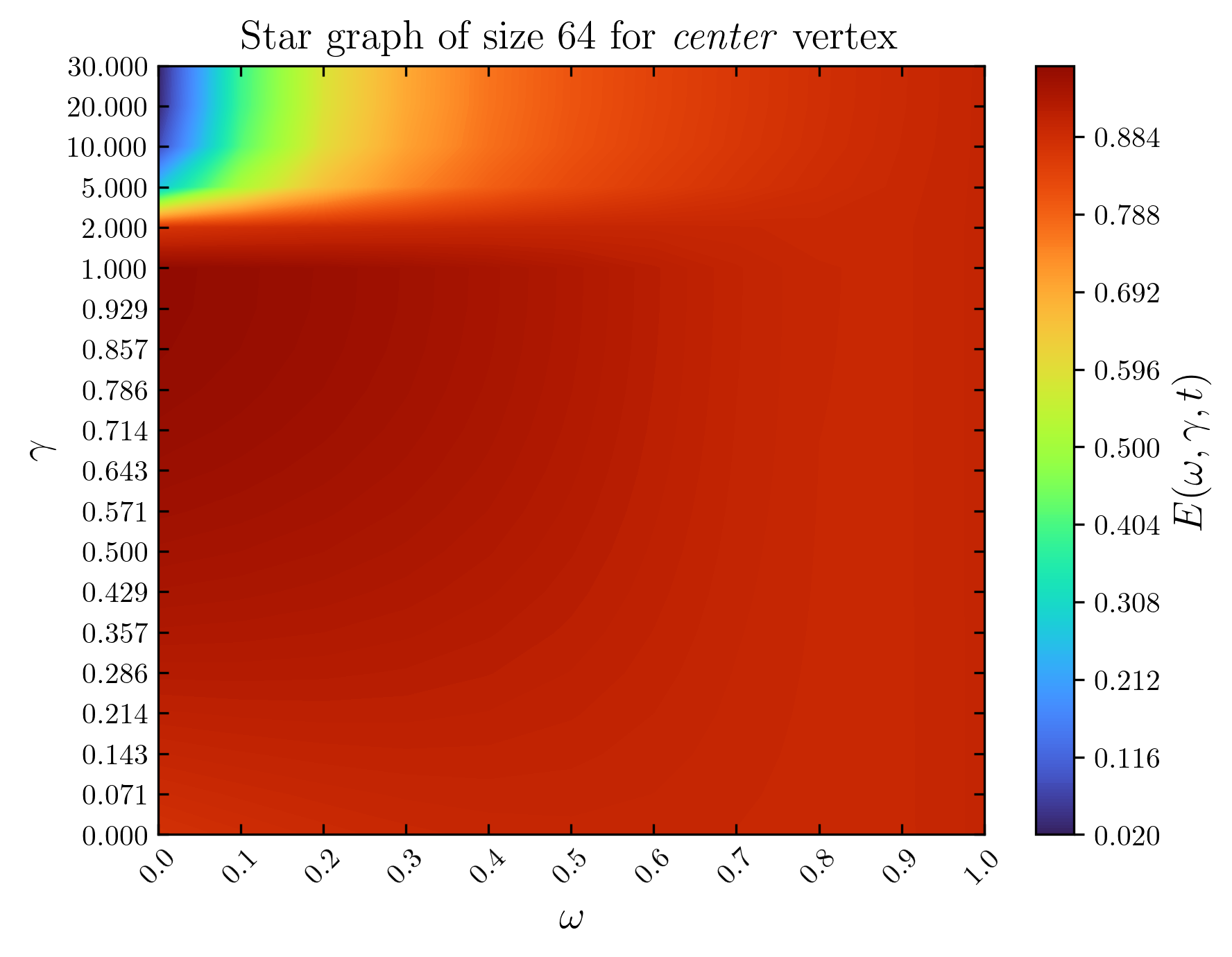}
    \includegraphics[width=0.3\textwidth]{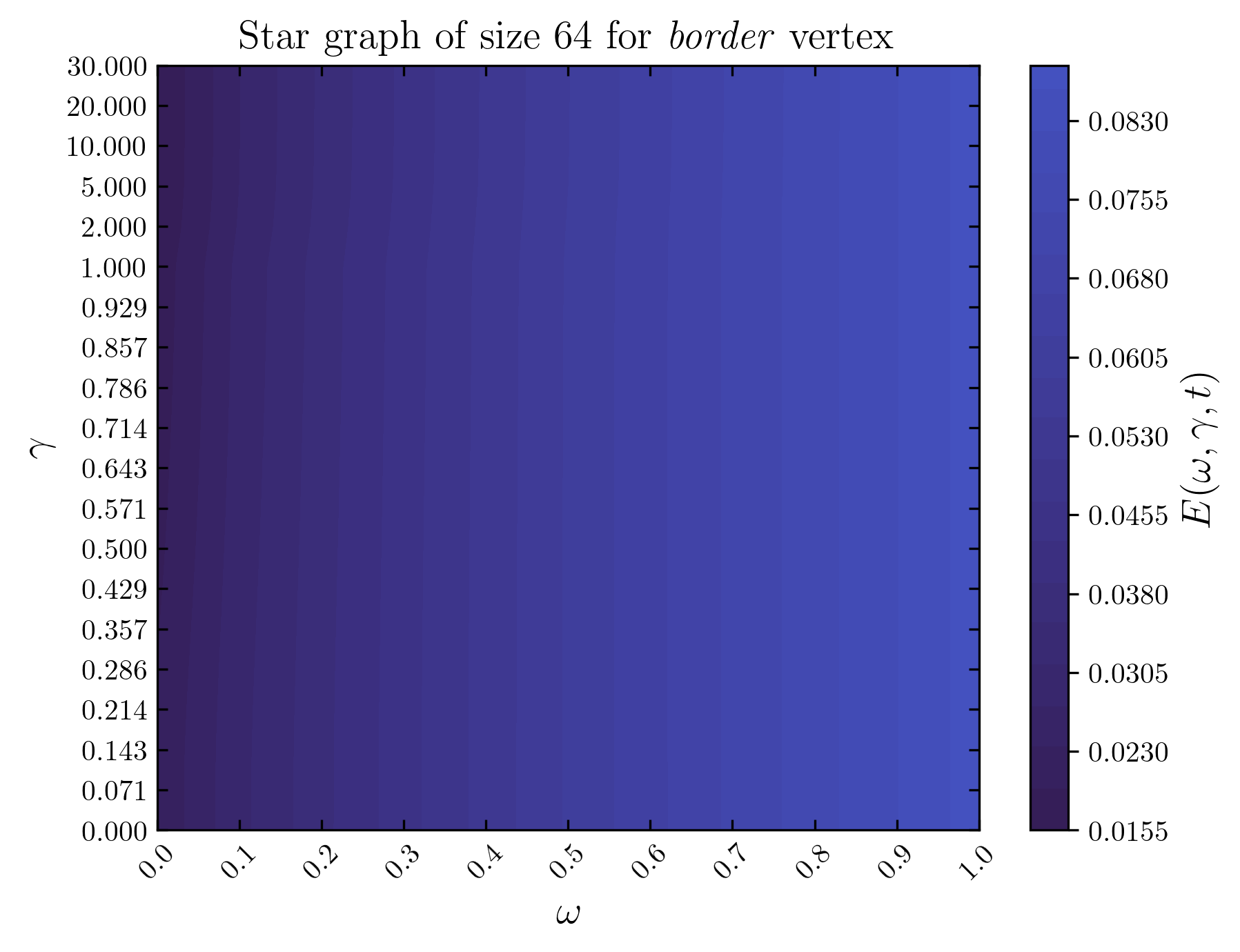}
    \includegraphics[width=0.3\textwidth]{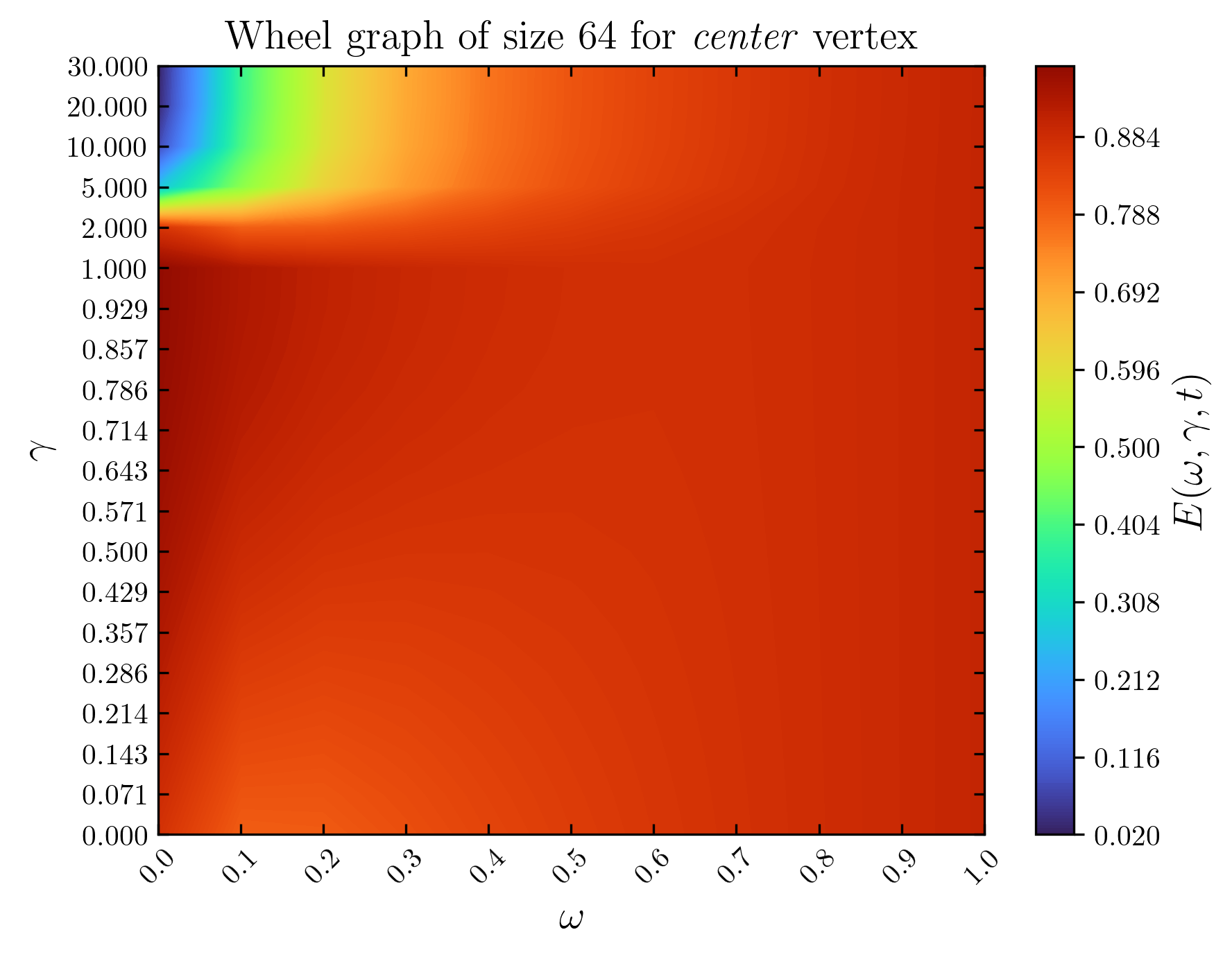}
    \includegraphics[width=0.3\textwidth]{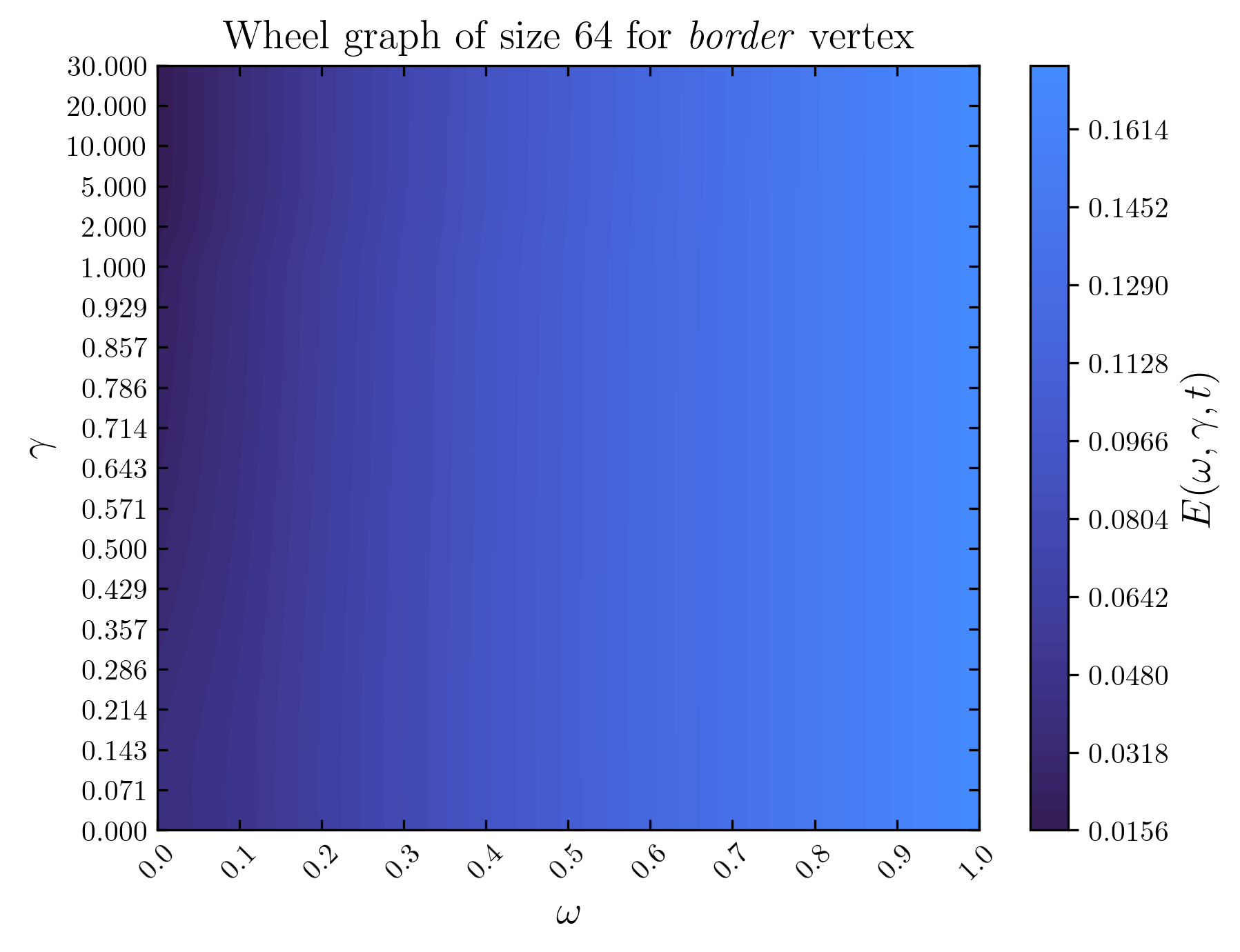}
    \includegraphics[width=0.3\textwidth]{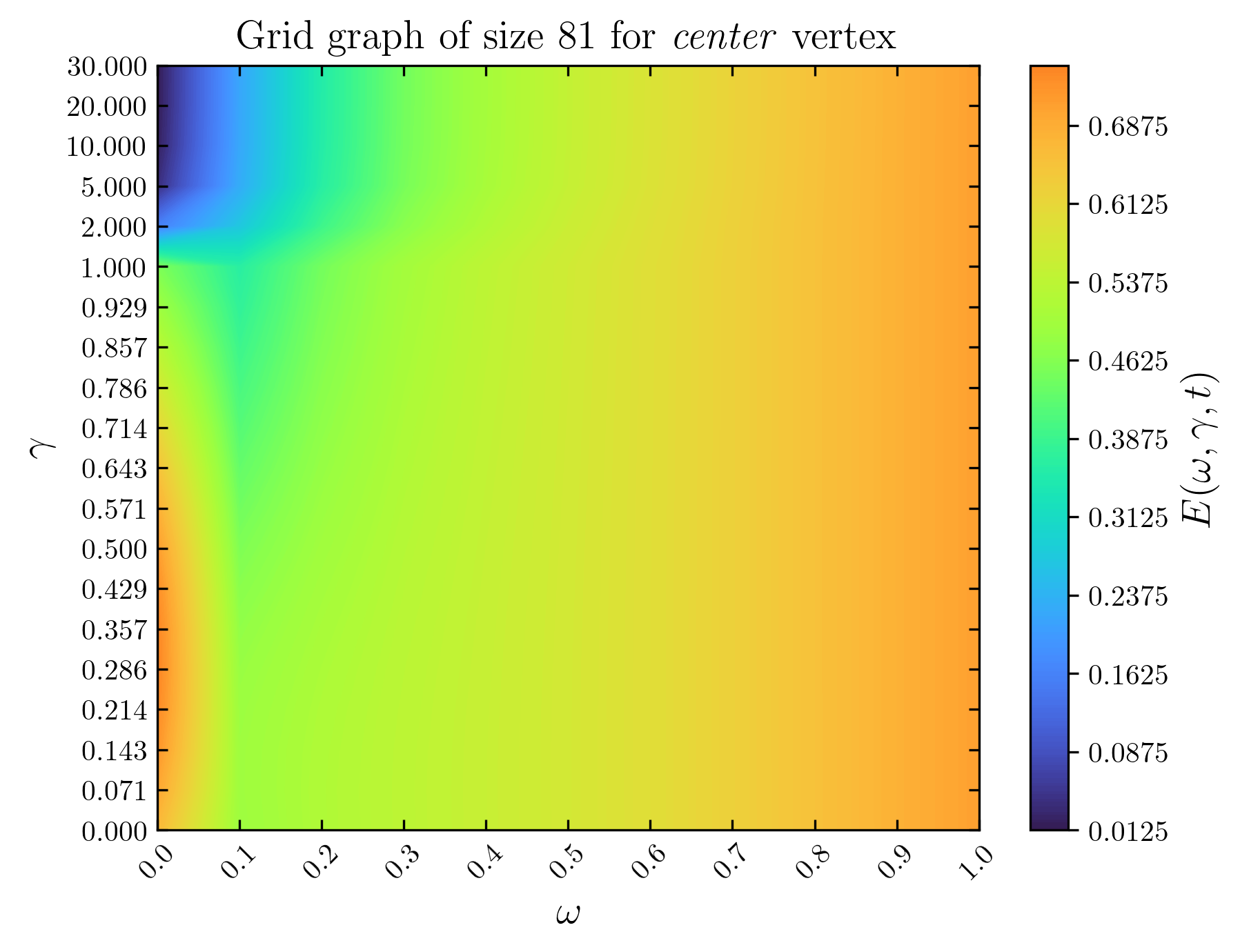}
    \includegraphics[width=0.3\textwidth]{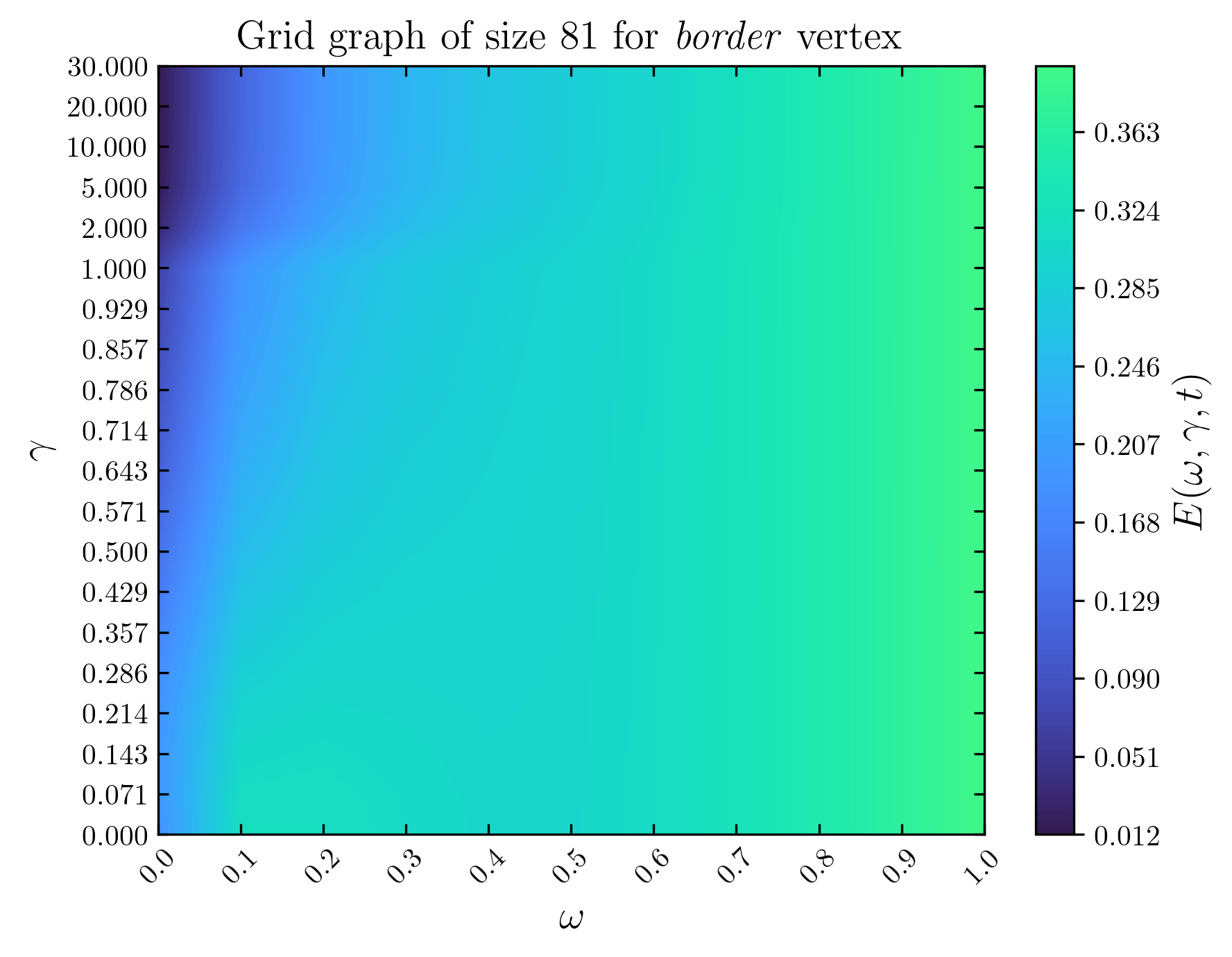}
    \includegraphics[width=0.3\textwidth]{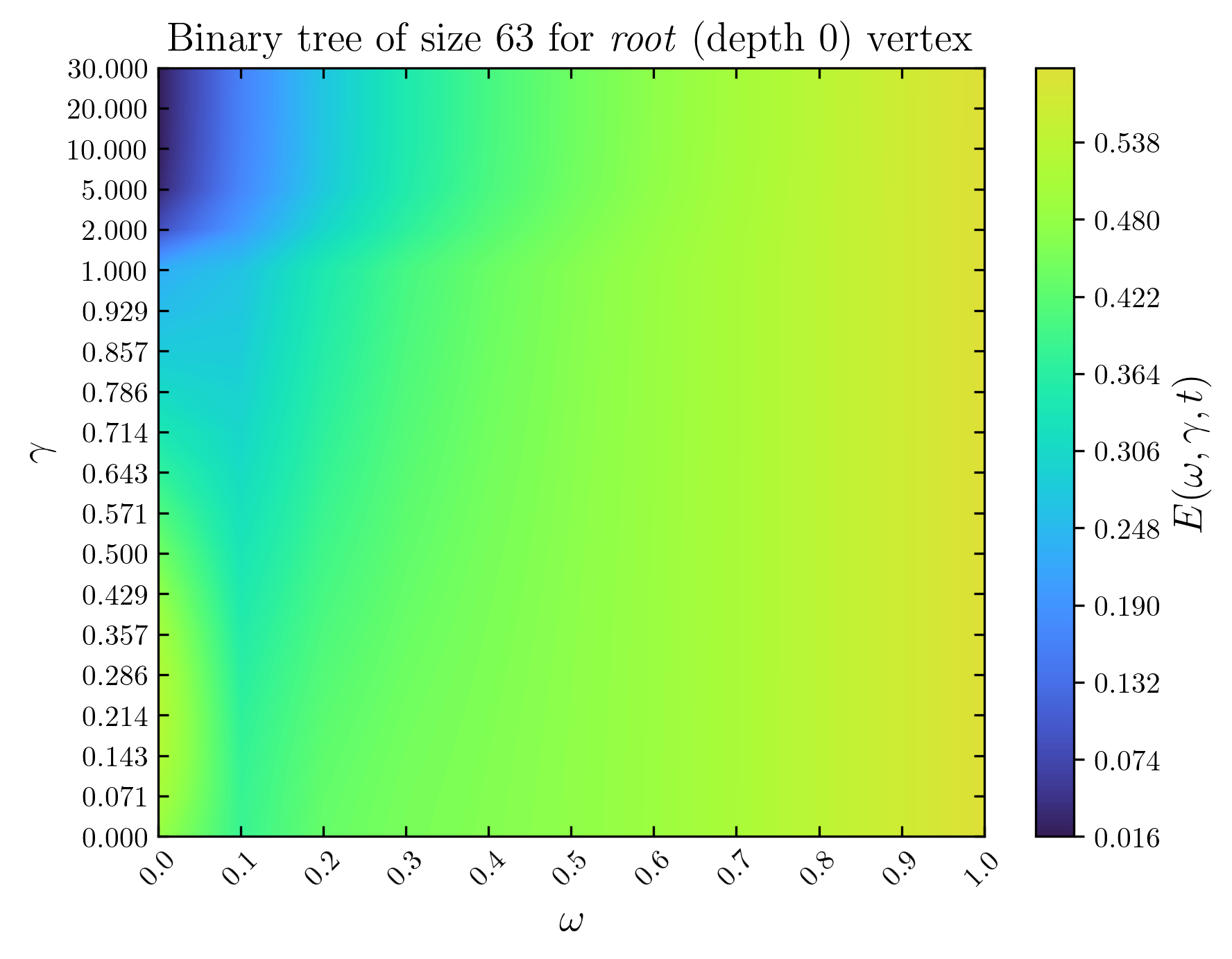}
    \includegraphics[width=0.3\textwidth]{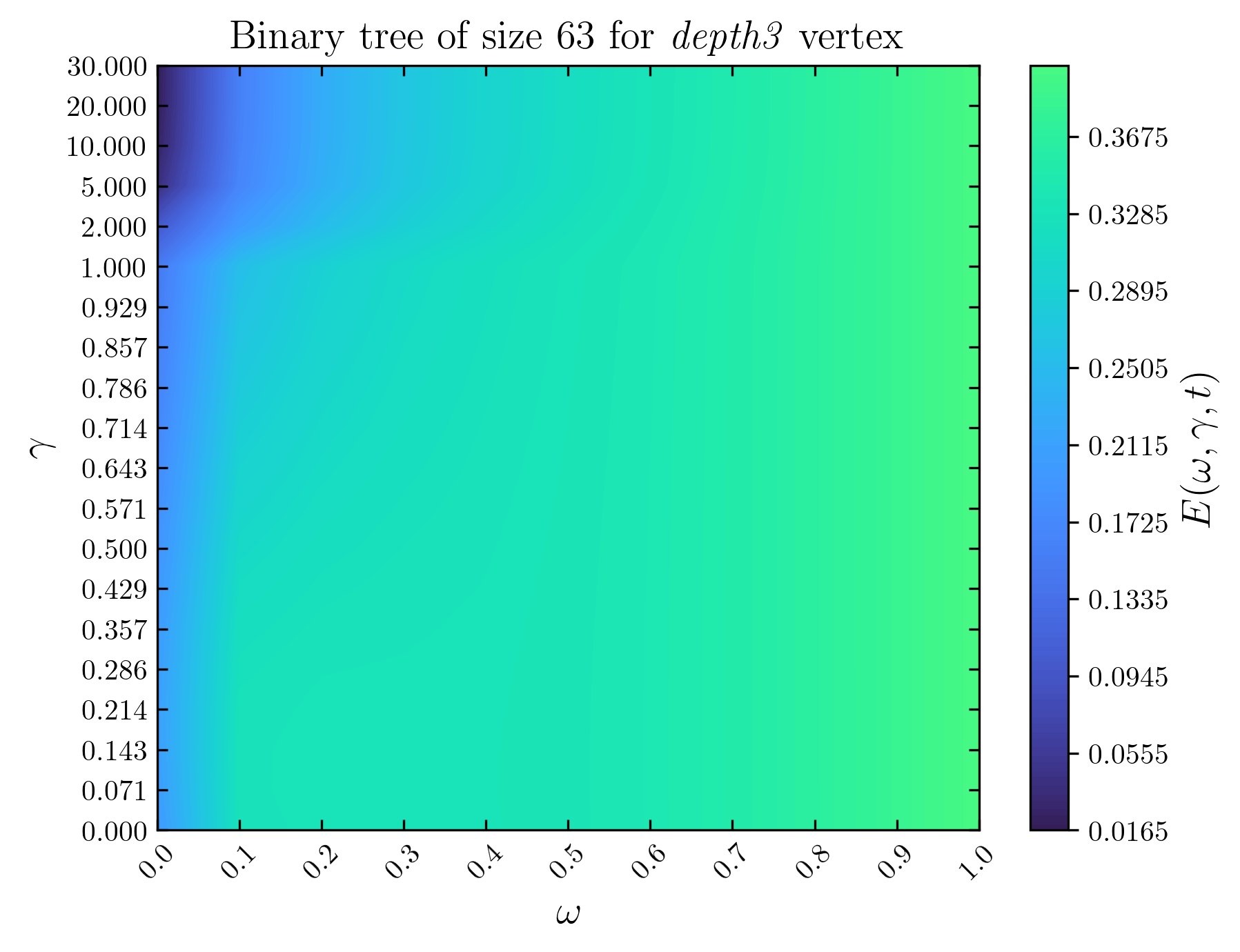}
    \includegraphics[width=0.3\textwidth]{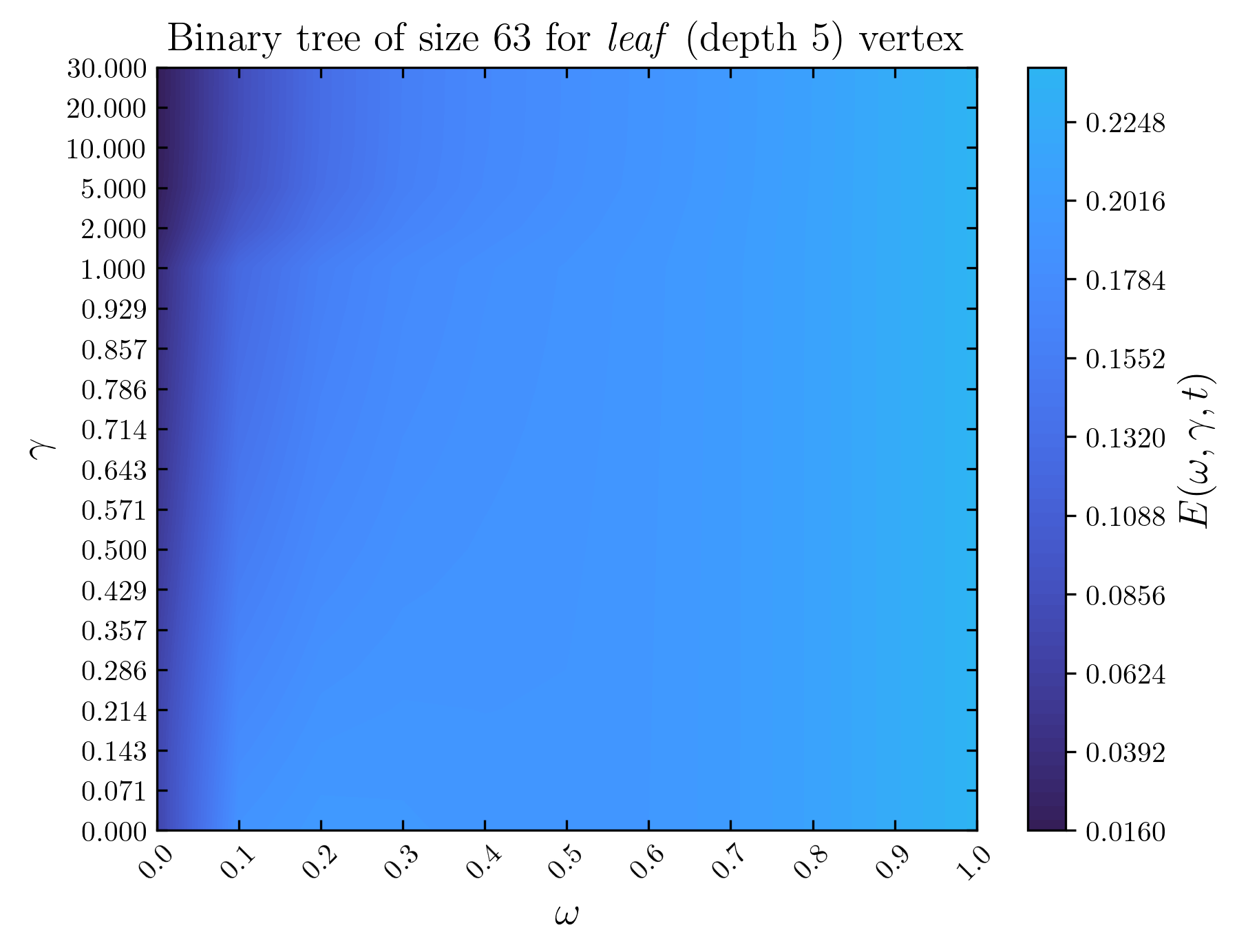}    
    \includegraphics[width=0.3\textwidth]{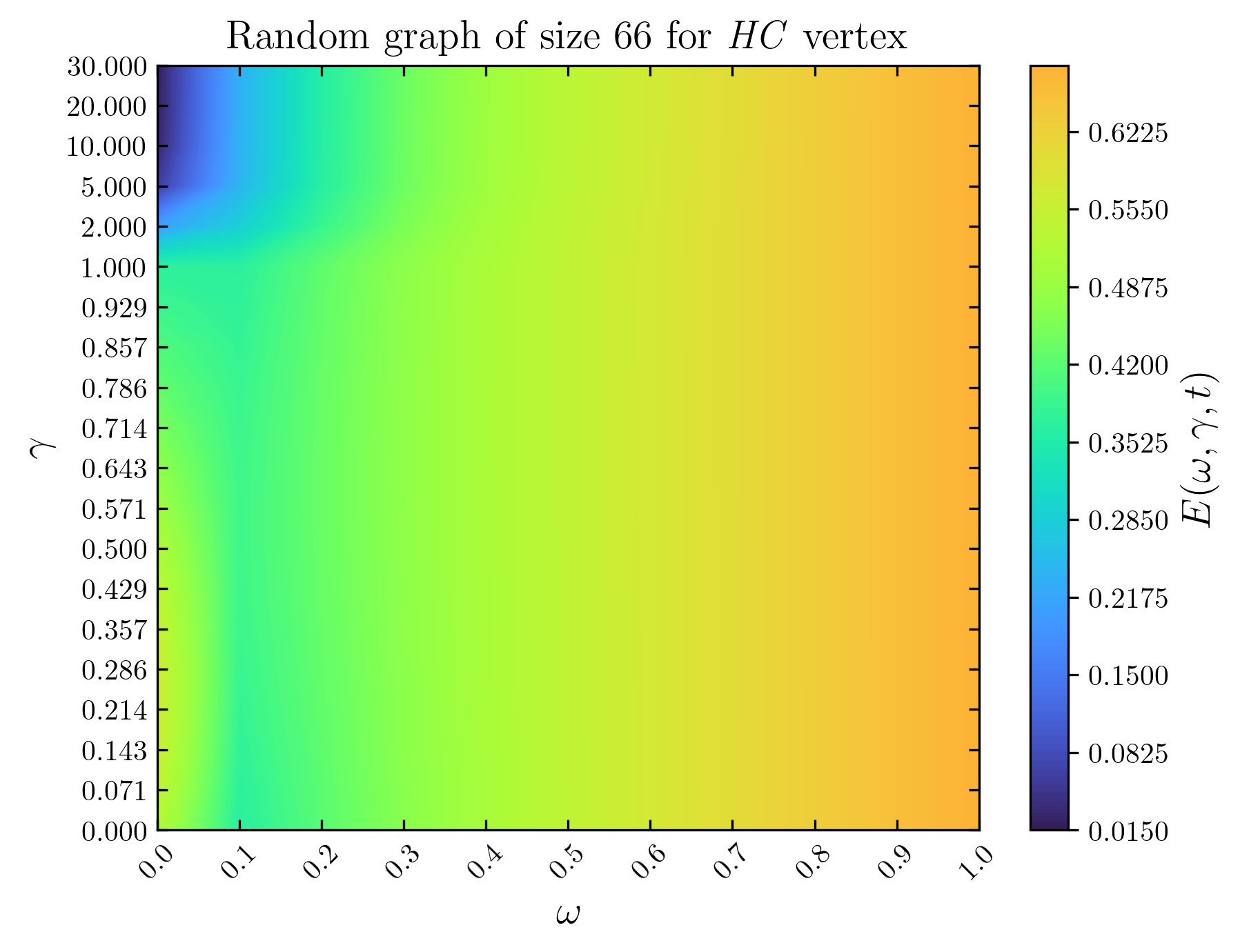}
    \includegraphics[width=0.3\textwidth]{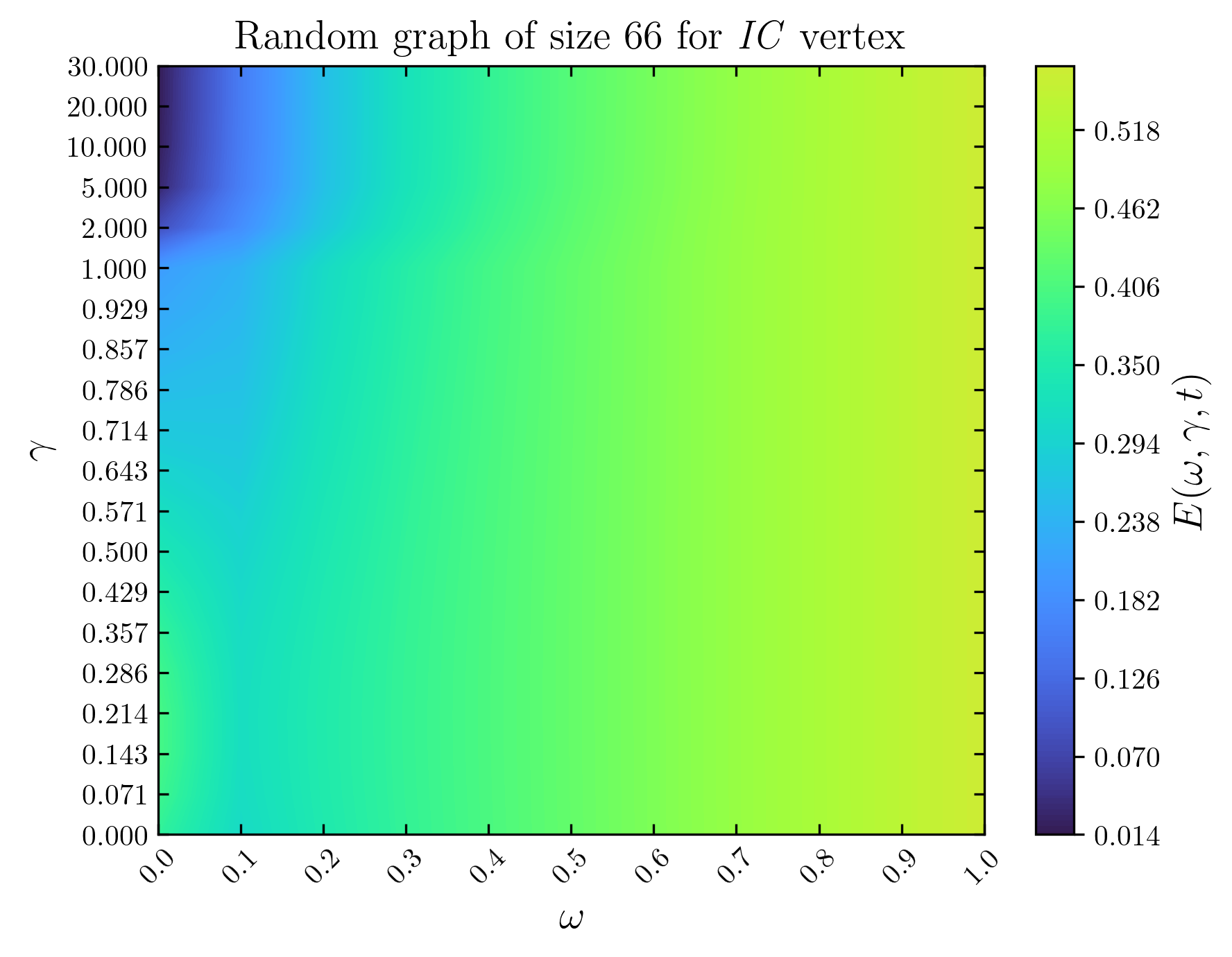}
    \includegraphics[width=0.3\textwidth]{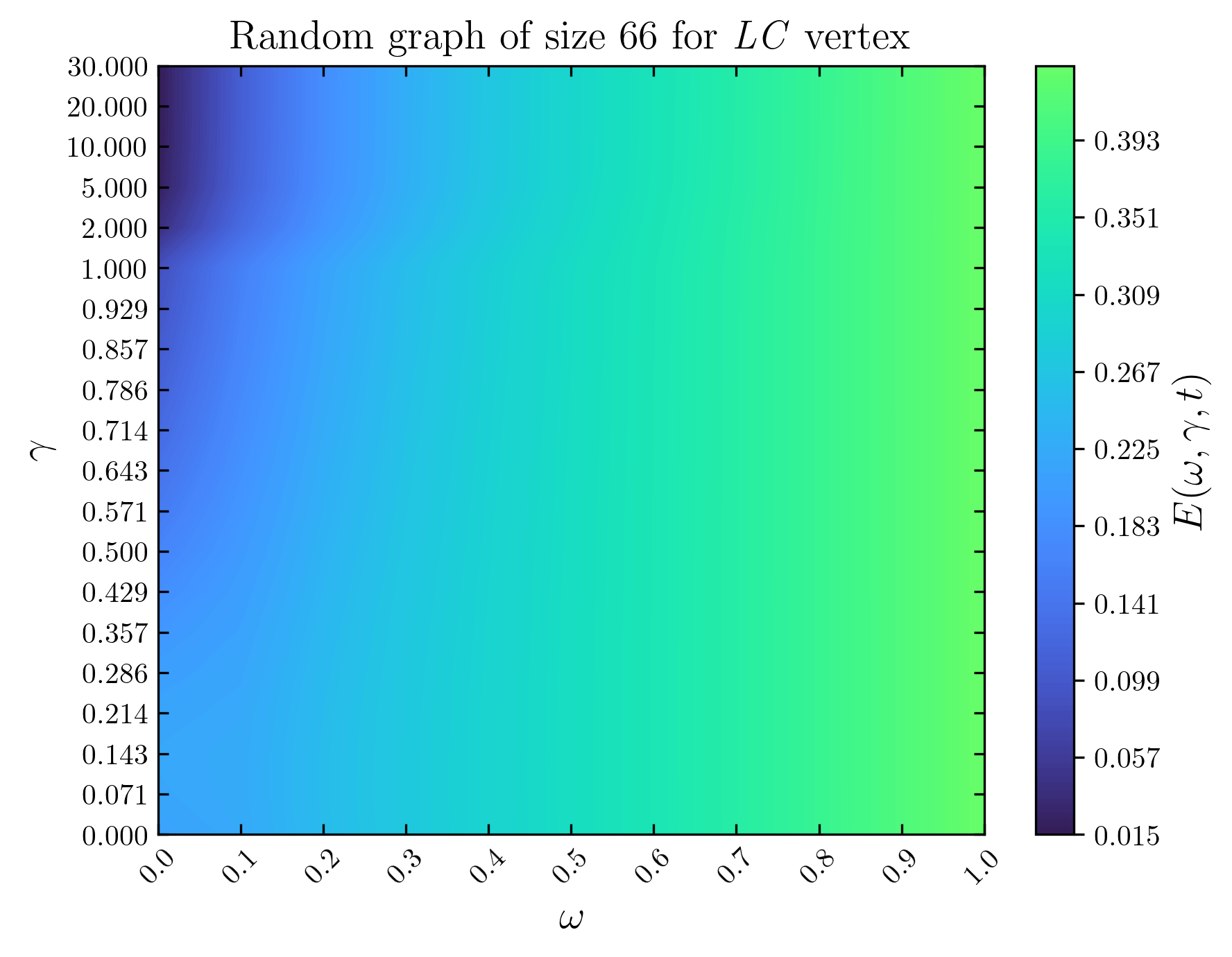}
    \caption{Stochastic Quantum Walk Search (SQWS) performances with different values of $\gamma\in [0,30]$ on the lollipop graph ($L_{32,32}$), the star graph ($S_{63}$), the wheel graph ($W_{64}$), the 2D-grid ($G_{9\times9}$), the perfect binary tree of depth 5 ($PBT_{5}$) and a random graph ($SW_{66}$) which was constructed by gluing together three small-world graphs of size $N=22$ each with different average connectivity and rewiring probabilities. The time evolution is set to $t=10N$ units of time, with $N$ the size of each graph. The quantum walk dynamics is recovered for $\omega=0$, the classical random walk for $\omega=1$, and a linear combination of the two when $\omega\in]0,1[$.}
    \label{fig:qsws_more}
\end{figure*}

\begin{figure}
    \centering
    \includegraphics[width=0.5\textwidth]{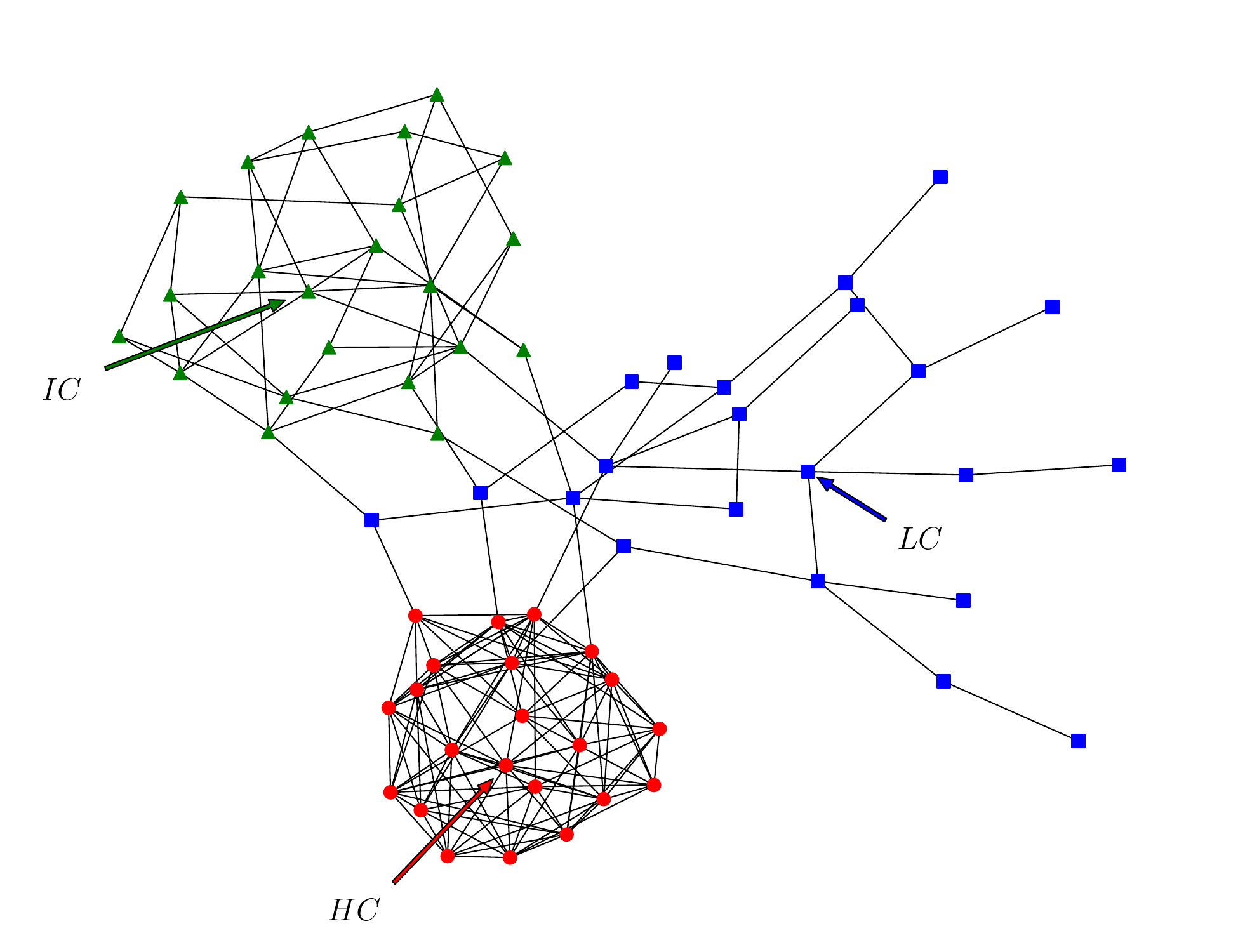}
    \caption{Random graph $SW_{66}$ constructed by gluing together three small-world graphs of size $N=22$ each with different average connectivity and rewiring probabilities. The target vertices are named \textit{HC}, \textit{IC} and \textit{LC}, respectively for high, intermediate and low connectivity. Their average connectivity are respectively 10, 4 and 3. Their respective rewiring probabilities are 0.1, 0.5 and 0.8.}
    \label{fig:random_graph}
\end{figure}

We now run the SQWS exclusively on non-vertex transitive graphs, i.e. graphs with a structure that distinguishes their vertices, of sizes $N\in[63,81]$. We select the lollipop graph ($L_{M,N}$), the star graph ($S_{N-1}$), the wheel graph ($W_{N}$), the 2D-grid ($G_{\sqrt{N}\times \sqrt{N}}$), the perfect binary tree of depth $d$ ($PBT_{d}$) and a random graph ($SW_{N}$) constructed by gluing together three small-world graphs of 22 vertices each with different average connectivity and rewiring probabilities~\cite{watts1998collective}. We show the graph $SW_{66}$ in Fig. \ref{fig:random_graph} and the results of the SQWS on these graphs in Fig. \ref{fig:qsws_more}. The lollipop consists of two graphs for which SQWS behaves differently, namely the complete graph and the path graph. Similar results are obtained for the \textit{complete} and \textit{shared} vertices with a transfer efficiency of 49\% for all values of $\omega\neq 0.1$ and when $\gamma<2$. There is a performance drop of about 10\% for the $\omega=0.1$ regime compared to the others. Finally, when $\gamma>2$, the quantum regime loses efficiency drastically, reaching 0.02\% for $\gamma=30$. When the target vertex is \textit{path}, however, the results are completely different: SQWS performance is extremely poor for all $\omega$ regimes and does not exceed 0.05\%. We also observe that, although low, the efficiency is insensitive to variations in $\gamma$. However, it increases slightly with the increase in $\omega$, from 0.01\% for $\omega=0$ to 0.05\% for $\omega=1$. For star and wheel graphs, the results are also similar to each other depending on the target vertex. When the target is the central vertex of the graph \textit{center}, we obtain respective efficiencies greater than 87\% and 80\% for all regimes $\omega$ when $\gamma\leq 2$. For both graphs, the quantum regime remains slightly more efficient than the others, with 97\% for $\gamma=1$. Furthermore, once $\gamma>2$, we also observe a decrease in performance, reaching 0.02\% for $\gamma=30$ for both graphs. We obtain very different results when the target vertex is \textit{border}, which is any vertex other than the central vertex. For both graphs, variations in $\gamma$ do not impact performance, and efficiency increases with increasing $\omega$. For the star graph, efficiency does not exceed 0.08\%, and 0.16\% for the wheel graph. Although efficiency is very low in both cases, the results for the wheel graph are twice as good as those for the star graph. For the grid, we mark the central vertex and a vertex located at one end. We obtain twice as good results for \textit{centre} as for \textit{border}. Furthermore, for \textit{centre}, the highest performance is obtained for $\omega=0$ and $\omega=1$, whereas for \textit{border}, it is for $\omega=1$. For the \textit{center} vertex, we observe that for the noisy regime, an increase in $\omega$ improves performance, reaching a maximum of around 70\% for $\omega=1$. As for the \textit{center} vertex, the quantum regime is never the most efficient; moreover, we observe that for low values of $\gamma\leq 0.3$, the low-noise regime $\omega=0.1$ performs better than the regimes $\omega\in[0.2,0.8]$, which we only observed for graphs with low density and centrality degree and high eccentricity. Of all the vertices on which we ran SQWS, the \textit{border} vertex of the grid most closely approximates these characteristics apart from those of the cycle, path, tadpole and maze, which explains the weak appearance of this phenomenon. For the perfect binary tree with depth 5, we mark three vertices located at different depths in the tree: \textit{root} (depth 0), a vertex \textit{depth3} (located at depth 3), and a leaf vertex at depth 5. Overall, we observe that the greater the depth of the target vertex, the poorer the performance of SQWS. Finally, for the random graph shown in Fig. \ref{fig:random_graph} composed of different graphs with different densities, we observe the same phenomenon, where performance decreases with the loss of density and connectivity of the target vertex.

\subsection{Ring-lattice transition}

\begin{figure}
    \centering
    \includegraphics[width=0.9\linewidth]{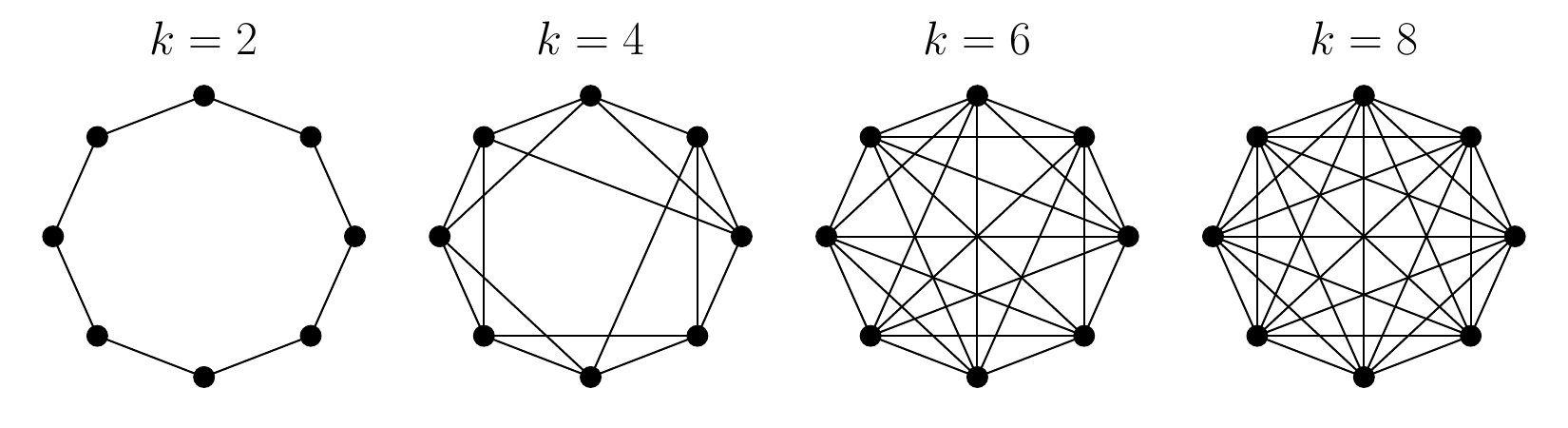}
    \caption{Transition from the cycle graph to the complete graph of size $N=8$ with the Ring-lattice model. As far as possible, each vertex is connected to its $k$-nearest neighbors.}
    \label{fig:ring_lattice}
\end{figure}

\begin{figure}
    \centering
    \includegraphics[width=0.9\linewidth]{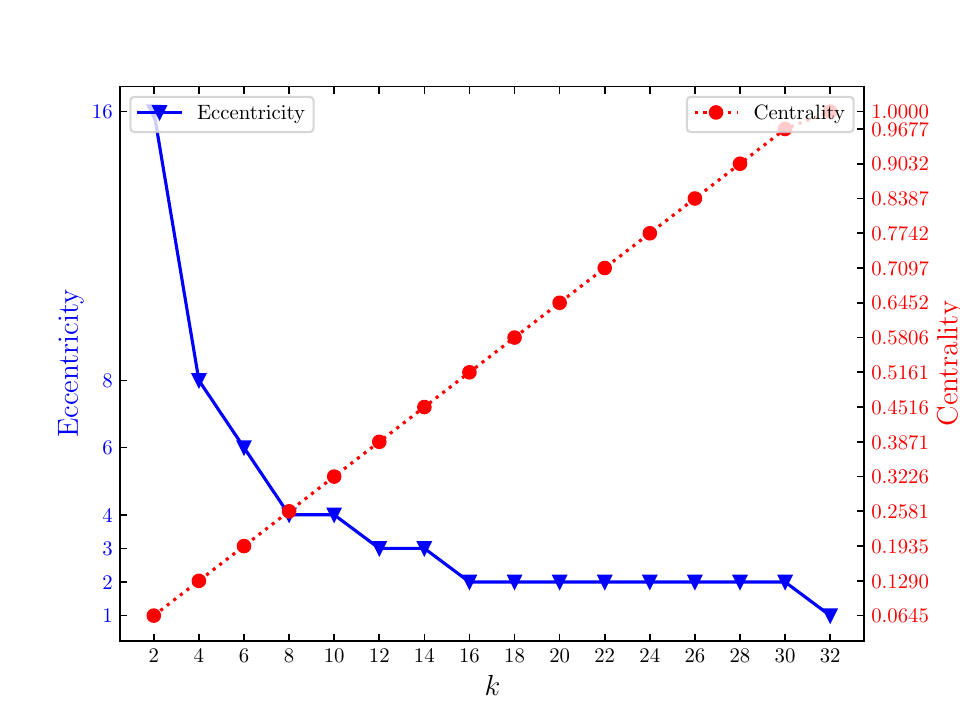}
    \caption{Eccentricity and centrality of the target vertex as a function of $k$ for Ring-lattice graphs of size $N=32$.}
    \label{fig:ring_lattice_change}
\end{figure}

\begin{figure*}
    \centering
    \includegraphics[width=0.8\linewidth]{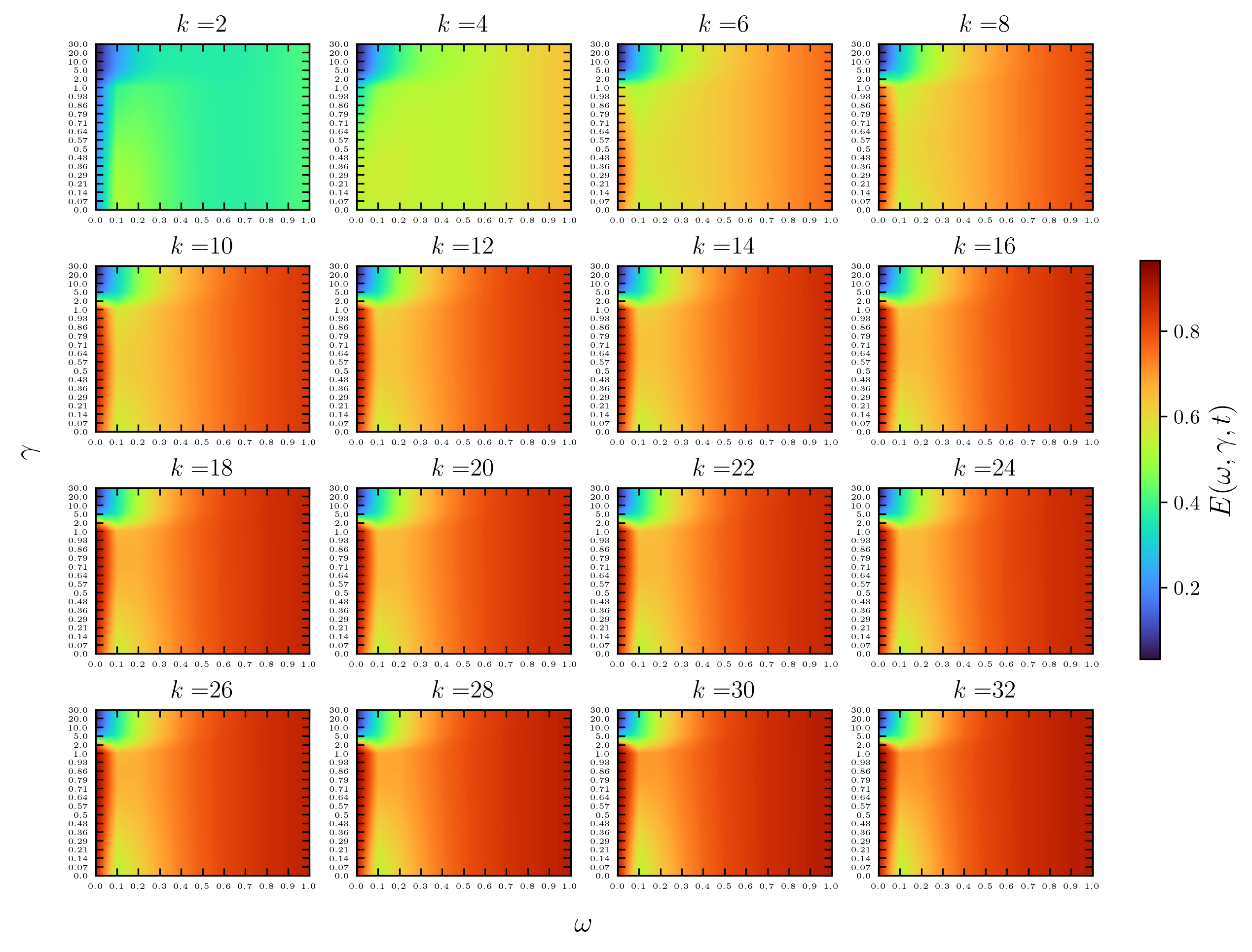}
    \caption{Stochastic Quantum Walk Search (SQWS) performances with different values of $\gamma\in [0,30]$ on Ring-lattice graphs of size $N=32$. The cycle graph is recovered for $k=2$ and the complete graph for $k=32$. An increase in the value of $k$ corresponds to increasing the average connection of each vertex with its $k$-nearest neighbors. The time evolution is set to $t=320$ units of time. The quantum walk dynamics is recovered for $\omega=0$, the classical random walk for $\omega=1$, and a linear combination of the two when $\omega\in]0,1[$.}
    \label{fig:qsws_ring_lattice}
\end{figure*}

As the results on the cycle and the complete graphs are completely different, we use the Ring-lattice graph model to progressively transform a cycle into a complete graph. A ring-lattice graph also known as a $k$-cycle is the basis of Watts and Strogatz model widely known as small-world networks~\cite{watts1998collective}. It consists of a cycle graph where each vertex is connected to its $k$ nearest neighbors. Therefore when $k=2$ we recover a cycle graph and the maximum value of $k$ generates a complete graph. As an illustration we show the transition from the cycle to the complete graph of size $N=8$ in Fig. \ref{fig:ring_lattice}. We run the SQWS on the Ring-lattice graph of size $N=32$ for 16 differents values of $k$, the cycle graph is obtained for $k=2$ and the complete graph for $k=32$. We present the results in Fig. \ref{fig:qsws_ring_lattice}. As $k$ increases, the eccentricity of the marked vertex decreases as the graph's connections increase, leading to an increase in its centrality as shown in Fig. \ref{fig:ring_lattice_change}. We observe that initially, with the graph cycle when $k=2$, the regime around $\omega=0.1$ is the most efficient of all, and the quantum regime does not perform well. Then, as the graph density increases, the other regimes perform better. Thus, from $k=6$ onwards, the $\omega=0.1$ regime is the least efficient for $\gamma<1$. With increased connectivity, we also see that the quantum regime supports higher values of $\gamma$ to maintain performance above 80\%. We note that the results seem to stabilise and are no longer sensitive to increases in connectivity from $k=10$ onwards, which corresponds roughly to the point at which the graph's eccentricity no longer really decreases with increases in average connectivity, as shown in Fig. \ref{fig:ring_lattice_change}.

\bibliographystyle{apsrev}
\bibliography{ref.bib}

\end{document}